\documentclass[acmsmall,nonacm]{acmart}

\AtBeginDocument{%
  \providecommand\BibTeX{{%
    \normalfont B\kern-0.5em{\scshape i\kern-0.25em b}\kern-0.8em\TeX}}}

\setcopyright{acmcopyright}
\usepackage{physics}
\usepackage{mathtools}
\usepackage{blkarray, bigstrut}
\usepackage{tikz}
\usetikzlibrary{quantikz}
\usepackage{caption}
\graphicspath{{./figures/}}
\usepackage{parskip}
\usepackage[ruled,vlined]{algorithm2e}
\usepackage{multicol}
\usepackage{graphicx}
\usepackage{javaLatexSTY/lstcustom}
\usepackage{float}
\usepackage{multirow}
\usepackage{todonotes}
\usepackage{balance}
\usepackage{soul}
\usepackage{enumitem}
\usepackage{caption}
\usepackage[shortcuts]{extdash}
\usepackage{subcaption}

\usepackage[normalem]{ulem}
\usepackage{listings}
\usepackage{lstcustom}
\usepackage[shortcuts]{extdash}
\newcommand*\circled[1]{\tikz[baseline=(char.base)]{
            \node[shape=circle,draw,inner sep=1pt] (char) {#1};}}

\begin{document}

\title{TornadoQSim: An Open-source High-Performance and Modular Quantum Circuit Simulation Framework}

\author{Ales Kubicek}
\authornote{The work presented in this paper is associated with his dissertation project at the University of Manchester.}
\email{akubicek@student.ethz.ch}
\affiliation{%
  \institution{ETH Zürich}
  \city{Zürich}
  \country{Switzerland}
}

\author{Athanasios Stratikopoulos}
\email{athanasios.stratikopoulos@manchester.ac.uk}
\affiliation{
  \department{Department of Computer Science}
  \institution{University of Manchester}
  \city{Manchester}
  \country{United Kingdom}
}

\author{Juan Fumero}
\email{juan.fumero@manchester.ac.uk}
\affiliation{
  \department{Department of Computer Science}
  \institution{University of Manchester}
  \city{Manchester}
  \country{United Kingdom}
}

\author{Nikos Foutris}
\email{nikos.foutris@manchester.ac.uk}
\affiliation{
  \department{Department of Computer Science}
  \institution{University of Manchester}
  \city{Manchester}
  \country{United Kingdom}
}

\author{Christos Kotselidis}
\email{christos.kotselidis@manchester.ac.uk}
\affiliation{
  \department{Department of Computer Science}
  \institution{University of Manchester}
  \city{Manchester}
  \country{United Kingdom}
}


\begin{abstract}
Quantum computers are driving a new computing paradigm to address important computational problems in science.
For example, quantum computing can be the solution to demystify complex mathematic formulas applied in cryptography, or complex models used in chemistry for biological systems.
Due to the early stage in the development of quantum hardware, simulation is currently playing a prime role in research.
To tackle the exponential cost of quantum simulation, state-of-the-art simulators are typically implemented using programming languages associated with High Performance Computing, 
while also providing the means for hardware acceleration on heterogeneous co-processors (e.g.,\ GPUs). 
The vast majority of quantum simulators implements a part of the simulator in a platform-specific language (e.g.,\ CUDA, OpenCL). 
This approach results in fragmented development as developers have to manually specialize the code for custom execution across different devices or microarchitectures.

In this article, we present TornadoQSim, an open-source quantum circuit simulation framework implemented in Java. 
The proposed framework has been designed to be modular and easily expandable for accommodating different user-defined simulation backends, such as the unitary matrix simulation technique.
Furthermore, TornadoQSim features the ability to interchange simulation backends that can simulate arbitrary quantum circuits.
Another novel aspect of TornadoQSim over other quantum simulators is the transparent hardware acceleration of the simulation backends on heterogeneous devices.
TornadoQSim employs TornadoVM to automatically compile parts of the simulation backends onto heterogeneous hardware, thereby addressing the fragmentation in development due to the low-level heterogeneous programming models.
The evaluation of TornadoQSim has shown that the transparent utilization of GPU hardware can result in up to 506.5$x$ performance speedup when compared to the vanilla Java code for a fully entangled quantum circuit of 11 qubits. 
Other evaluated quantum algorithms have been the Deutsch-Jozsa algorithm (493.10$x$ speedup for a 11-qubit circuit) and the quantum Fourier transform algorithm (518.12$x$ speedup for a 11-qubit circuit).
Finally, the best TornadoQSim implementation of unitary matrix has been evaluated against a semantically equivalent simulation via Qiskit.
The comparative evaluation has shown that the simulation with TornadoQSim is faster for small circuits, while for large circuits Qiskit outperforms TornadoQSim by an order of magnitude.
\end{abstract}




\maketitle

\section{Introduction}
Quantum computing is an intersection across the field of physics, mathematics, electrical engineering and computer science.
Since the first theoretical notion in early eighties, quantum computing has driven a new computing paradigm that can leverage quantum mechanical properties as a means to solve complex problems that is not feasible to be solved by a conventional computer~\cite{Hirvensalo2013}.
For example, quantum computing can become the solution for decoding complex mathematic formulas applied in cryptography ~\cite{DBLP:journals/corr/abs-1804-00200}, or complex models used in chemistry for biological systems~\cite{Hirvensalo2014} and machine learning ~\cite{Biamonte_2017, doi:10.1080/00107514.2014.964942}.
A universal theoretical model of a quantum computer has been proposed~\cite{1985RSPSA.400...97D}, numerous quantum algorithms have been developed to solve some of the practical problems, and a limited quantum hardware has been built to demonstrate the feasibility of the physical implementation. 
As a matter of fact, multiple important milestones have been achieved in demonstrating quantum advantage over classical computing. This includes Google's demonstration of quantum supremacy for random-sampling in 2019~\cite{arute2019} as well as Xanadu's Gaussian boson sampling in 2022~\cite{madsen2022quantum}.
Due to the high complexity in building quantum hardware, the majority of recent research has focused on quantum simulation, as a way to study the design and implementation of a quantum computer~\cite{doi:10.1126/science.273.5278.1073}.

Quantum simulators are engineered to facilitate physicists and mathematicians for the creation of custom quantum circuits tailored to specific problems.
The state-of-the-art quantum simulators (e.g.,\ Qiskit~\cite{qiskit_docs2018, qiskit_docs2021}, Cirq~\cite{cirq_docs2021}, QDK~\cite{qdk_docs2021}, QuEST ~\cite{quest2019}, qFlex ~\cite{qflex2019}, IQS ~\cite{iqs2020}, ProjectQ ~\cite{projectq2018}) are built in C++, while also offering an interface to Python.
Additionally, quantum simulators exploit heterogeneous hardware, such as GPUs~\cite{heng2020} and FPGAs~\cite{lee2016, pilch2019}, as a means to accelerate part of the simulation.
However, the developers of these simulators are required to write the accelerated code in low-level programming models, such as CUDA~\cite{cuda_spec2021} and OpenCL~\cite{opencl2010}, or invoke pre-built kernels.
Although, quantum simulators have been available for C++ developers, there is limited availability of simulators built in managed programming languages (e.g.,\ Java) that ease code maintainance.
In addition, this is further exacerbated by the inability of Java to harness hardware acceleration.

This paper aims to alleviate the gap between slow performance and high readability of quantum simulators written in high-level managed programming languages, such as Java. 
In particular, we propose a modular architecture of a simulation framework that can be easily expanded to accommodate different user-defined simulation backends.
Additionally the proposed architecture employs hardware acceleration to reduce the simulation time, by employing the TornadoVM JIT compiler~\cite{tornado2018, tornado2019, Papadimitriou_2020} to generate GPU kernels in a seamless manner to the users.
The proposed system can be used as an educational and practical framework for programmers who aim to model quantum circuits from Java while also harnessing the performance benefits of hardware acceleration.
In a nutshell, it makes the following contributions:
\begin{itemize}
    \item It presents TornadoQSim, an open-source quantum circuit simulation framework implemented in Java. The code is available in GitHub\footnote{\url{https://github.com/beehive-lab/TornadoQSim}} under the Apache 2.0 license.
    The proposed framework has been designed to be modular and easily expandable for accommodating different simulation backends, such as the unitary matrix backend.
    It outlines the ability to interchange simulation backends that can simulate arbitrary quantum circuits.
    \item It describes the novel aspect of TornadoQSim over other quantum simulators, which is the transparent hardware acceleration of the simulation backends on heterogeneous devices.
    TornadoQSim employs TornadoVM to automatically compile parts of the simulation backends written in Java onto heterogeneous hardware, thereby addressing the fragmentation in development caused by the low-level heterogeneous programming models (e.g., OpenCL, CUDA).
    \item Finally, it evaluates the performance of the vanilla Java implementation of TornadoQSim against the hardware accelerated simulations, showcasing that the transparent utilization of GPU hardware can result in speedup up to 506.5$x$, 493.10$x$ and 518.12$x$ for a fully entangled circuit, a Deutsch-Jozsa quantum algorithm, and a quantum Fourier transform algorithm, respectively.
    Furthermore, it evaluates the best TornadoQSim implementation of the unitary matrix simulation backend against a semantically equivalent simulation via Qiskit, demonstrating that the simulation with TornadoQSim is faster for small circuits, while for larger circuits Qiskit outperforms TornadoQSim by an order of magnitude.
\end{itemize}
\section{Background in Quantum Systems}
This section outlines the foundations in quantum computing to provide the means for understanding how quantum systems are being built.
In particular, Section ~\ref{sec:key_principles} presents the basic principles of quantum theory, while Section~\ref{sec:data_representation} presents the main computational unit used in quantum computers.
Section~\ref{sec:quantum_gates} describes the logic behind quantum gates.
Finally, Sections~\ref{sec:quantum_cirquits}, ~\ref{sec:quantum_programs} and ~\ref{sec:quantum_simulation_backends} outline the theory behind quantum circuits, quantum programs as well as the simulation backends used for mathematically calculating quantum-level problems.
% Finally, Section ~\ref{sec:hardware_acceleration} highlights the importance of hardware acceleration in state-of-the-art quantum simulators.

\subsection{Key Principles in Quantum Computing}
\label{sec:key_principles}
To understand the basics in quantum computing, it is necessary to give an overview of some key quantum mechanical principles. 
At first, \textit{quantum state} is a term used to describe the behaviour of a particle (e.g.,\ electron, neutron) in a quantum system.
A state is typically represented by a vector in a complex vector space, or it can be equivalently expressed with the bra-ket notation as $\ket{\psi}$.
Figure~\ref{fig:hydrogen_atom} presents a quantum system that comprises of a hydrogen atom with a single electron and two orbitals.
In this system, the electron can collapse into two finite states: (i) the state in which the electron is in the ground state (lower orbital); and (ii) the state in which the electron is in the first excited state (upper orbital)~\cite{imperial_notes2015}.
Therefore, the system presented in Figure ~\ref{fig:hydrogen_atom} has two states, as follows:

\begin{figure}[htbp!] 
    \centering    
    \includegraphics[width=0.4\textwidth]{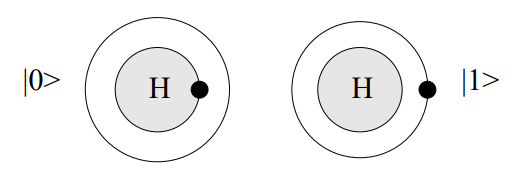}
    \caption{Hydrogen atom as a two-state Quantum system (Adapted from \cite{gruska1999}).}
    \label{fig:hydrogen_atom}
\end{figure}

%\textit{Bra} \bra{\phi}} defines a linear map in a complex vector space, while \ket{\psi}} defines a vector in a complex vector space. 
%In addition, the notation can be combined as ⟨φ|ψ⟩ to define a linear map \bra{\phi}} acting on a vector \ket{\psi}}, or in other words the inner product of the two vectors~\cite{mit_notes2013}.

% \newline
% \noindent\begin{minipage}{.5\linewidth}
% \[
%     \ket{0} = 
%         \begin{bmatrix}
%             1 \\ 
%             0
%         \end{bmatrix}
% \]
% \end{minipage}%
% \begin{minipage}{.5\linewidth}
% \[
%     \ket{1} = 
%         \begin{bmatrix}
%             0 \\ 
%             1
%         \end{bmatrix}
% \]
% \end{minipage}
% \newline

\subsubsection{Superposition}
Considering the quantum system presented in Figure~\ref{fig:hydrogen_atom}, the electron, when observed, will collapse into one of the two finite states ~\cite{gruska1999, imperial_notes2015}. Thus, the only information that can be physically observed is a binary value (0 or 1).
However, in practice, there is an infinite number of combined states that the particle can reach before collapsing.
This phenomenon is described as \textit{superposition} ~\cite{gruska1999}.
In superposition, the state is formulated as the aggregation of the probability amplitudes of the first and second state.
Equation~\ref{eq1} presents the formula that describes the quantum state using vectors, where $\alpha$ and $\beta$ are complex probability amplitudes; \(\alpha, \beta \in \mathbb{C}\) and \(\ket{\psi} \in \mathbb{C}^2\).
In addition, the probability amplitudes must be normalized, as shown in Equation~\ref{eq2}; in which $\abs{\alpha}^2$ and $\abs{\beta}^2$ express the probability of the electron being in the ground state and the first excited state, respectively.
In essence, the electron lives in a superposition of the final two states of the system.
The semantically equivalent formula is expressed with the bra-ket notation in Equation~\ref{eq3}.

\begin{equation}\label{eq1}
    \ket{\psi}=\alpha
    \begin{bmatrix}
            1 \\ 
            0
    \end{bmatrix}
    +\beta
    \begin{bmatrix}
            0 \\ 
            1
    \end{bmatrix}
    =
    \begin{bmatrix}
            \alpha \\ 
            \beta
    \end{bmatrix}
\end{equation}

\begin{equation}\label{eq2}
    1 = \abs{\alpha}^2 + \abs{\beta}^2
\end{equation}

\begin{equation}\label{eq3}
    \ket{\psi}=\alpha\ket{0}+\beta\ket{1}
\end{equation}

\subsubsection{Entanglement}
Another phenomenon that is known in quantum theory is \textit{entanglement}, which regards the binding between multiple particles for the state of a quantum system.
Considering the example of two particles, they can be transformed into a Bell state; also known as EPR state.
In this state, an internal bond is created that is independent of the spatial position of the particles.
%One way to describe such state is:
%\begin{equation}
%    \ket{\psi} = \frac{1}{\sqrt{2}}\ket{00} + 0\ket{01} + 0\ket{10} + \frac{1}{\sqrt{2}}\ket{11}
%\end{equation}
For example, when the first particle is measured, the state 0 is observed with a probability $\frac{1}{2}$.
Therefore, the quantum state of each particle collapses into 0 or into 1.
However, once the state of one particle is collapsed and the quantum state is re-normalized, the state of the second particle is no longer independent and will be observed with a probability of $1$ based on the result of the previous observation.
This phenomenon is also valid when particles are separated by a large spatial distance.
In 2017, Ren \emph{et al.}~\cite{ren2017} demonstrated an experiment that validated this phenomenon by using photons over a distance up to 1400 km.

\subsection{Quantum Bits \& Quantum Registers}
\label{sec:data_representation}
\begin{figure}[htbp!] 
    \centering    
    \includegraphics[width=0.5\textwidth]{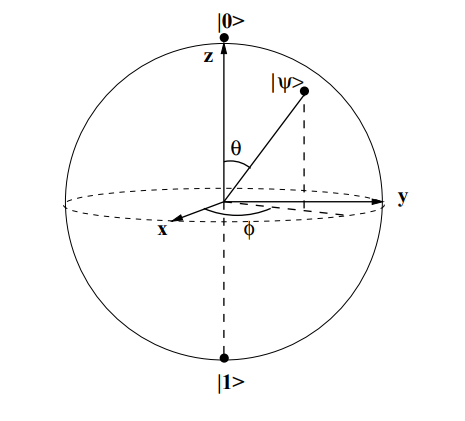}
    \caption{The Bloch sphere (Adapted from \cite{imperial_notes2015}).}
    \label{fig:bloch}
\end{figure}

% Several important questions emerge when designing a computational device. For instance, \emph{how is information represented, stored and manipulated to solve a computational problem?}
% In a conventional computer, a \texttt{bit} is the smallest unit of information with two discrete states 0 and 1. 
% Multiple bits can compose larger units, such as a \texttt{byte}, that consists of eight bits. 
To process the information stored in a bit, a conventional processor must first load it in a register;
then, perform an arithmetic or logic operation over the loaded data; and finally, store the result back either in a register or in a memory address.
Similar concepts are applied by quantum computers.
The basic unit for storing information in a quantum computer is a quantum bit or qubit ~\cite{Hirvensalo2013, gruska1999, imperial_notes2015}.
The qubit is a two-state quantum system with a computational basis {$\ket{0}$, $\ket{1}$}, as expressed in Equation~\ref{eq3}. 
This mathematical definition of a qubit can be also represented in a form of the Bloch sphere~\cite{imperial_notes2015}, as shown by Figure~\ref{fig:bloch}.
The top pole of the sphere represents the state $\ket{0}$, while the bottom pole of the sphere represents the state $\ket{1}$.
The quantum state $\ket{\psi}$ of the system is a point anywhere on the surface of the sphere.
The spherical representation of the qubit state motivates the need for quantum gates (e.g.,\ phase gates, etc.) that enable every point on the sphere to be reached from the initial state $\ket{0}$ or $\ket{1}$.

Additionally, a quantum system can comprise more than one (\textit{n}) qubits.
In this case, the number of required complex amplitudes to define the quantum state is $2^n$ and $\ket{\psi} \in {\mathbb{C}^2}^n$, thereby growing exponentially with the number of qubits. Entities composed of multiple qubits are called quantum registers.

\subsection{Quantum Gates}
\label{sec:quantum_gates}
Quantum gates are transformations that are semantically equivalent with the digital logic gates in digital circuits.
Quantum gates operate on qubits, and instead of a truth table, the quantum gates are defined by unitary linear maps \cite{imperial_notes2015}. 
The reason why a linear map must be unitary is to preserve the normalized quantum state after the operation is performed \cite{Hirvensalo2013, gruska1999}.
For a linear map to be unitary, it means to be defined by a unitary matrix.
Any $n \times n$ matrix $U$ is unitary, if and only if, the adjoint of $U$ is equal to the inverse of $U$, or mathematically $U^\dagger=U^{-1}$.

Table~\ref{table:quantumgates} presents the standard single qubit quantum gates along with their unitary matrices.
Phase gates, such as Pauli-X, Y and Z gates, are defined by unitary matrices of the same name, and they flip the quantum state by 180\textdegree~ around x, y and z axis, respectively.
Moreover, the Hadamard gate \cite{imperial_notes2015} allows the tranformation of a qubit into a superposition and the phase shift quantum gate R can be used to arbitrarily control the phase of the qubit.
% A NAND logic gate is universal in the classical setting, which means that any logic gate can be constructed from a combination of NAND gates. 
The Hadamard gate, the phase gates (Pauli-X, Y, Z and phase shift R) and CNOT gate compose a set of quantum gates that can be combined to construct any computable $n$-qubit unitary operation~\cite{imperial_notes2015}.
% Moreover, even though NAND gate is not reversible, it can be implemented by using a quantum gate with additional qubit to preserve the input information.
% Specifically, three qubit Toffoli gate implements the logic of the classical two input NAND gate.
% This fact proves that any classical computer can be simulated by a quantum computer \cite{imperial_notes2015}.

\begin{table}[htbp!]
    \centering
    \begin{tabular}{l c}
    \toprule
        Quantum Gate & Unitary Matrix  \\  
    \midrule
        $H$ (Hadamard) & 
        $
        \begin{bmatrix}
            \frac{1}{\sqrt{2}} & \frac{1}{\sqrt{2}} \\ 
            \frac{1}{\sqrt{2}} & \frac{-1}{\sqrt{2}}
        \end{bmatrix} 
        $ \\
        
    \midrule
        $X$ (Pauli-X) & 
        $
        \begin{bmatrix}
            0 & 1 \\ 
            1 & 0
        \end{bmatrix} 
        $  \\
    \midrule
        $Y$ (Pauli-Y) & 
        $ \begin{bmatrix}
            0 & -i \\ 
            i & 0
        \end{bmatrix} $ 
        \\
    \midrule
        $Z$ (Pauli-Z) & 
        $ \begin{bmatrix}
            1 & 0 \\ 
            0 & -1
        \end{bmatrix} $ 
        \\
    \midrule
        $R_\phi$ (Phase Shift) & 
        $
        \begin{bmatrix}
            1 & 0 \\ 
            0 & e^{i\phi}
        \end{bmatrix}
        $  \\
    \bottomrule 
    \end{tabular}
    \vspace{0.3cm}
    \caption{Single qubit quantum gates.}
    \label{table:quantumgates}
\end{table}

Equation~\ref{eq4} presents the mapping of qubit states for the Pauli-X gate (i.e.,\ the NOT operation in classical setting).
Furthermore, Equation~\ref{eq5} shows the unitary matrix for the same gate.
The application of that quantum gate on a single qubit system can be mathematically expressed as shown in Equation~\ref{eq6}.
The probability amplitudes after the application of the quantum gate will be swapped.
Thus, the qubit will be inverted, which is the expected outcome of the classical NOT operation.

\begin{equation}\label{eq4}
\begin{split}
    \ket{0} \mapsto \ket{1} \\
    \ket{1} \mapsto \ket{0}
\end{split}
\end{equation}

\begin{equation}\label{eq5}
    X = 
    \begin{blockarray}{*{2}{c} l}
        % \begin{block}{*{5}{>{$\footnotesize}c<{$}} l}
        %   \ket{0} & \ket{1} \\
        % \end{block}
        \begin{block}{[*{2}{c}]>{$\footnotesize}l<{$}}
          0 & 1 \\
          1 & 0 \\
        \end{block}
    \end{blockarray}
\end{equation}

\begin{equation}\label{eq6}
    X\cdot\ket{\psi} = 
    \begin{bmatrix}
        0 & 1 \\
        1 & 0
    \end{bmatrix}
    \cdot
    \begin{bmatrix}
        \alpha \\
        \beta
    \end{bmatrix}
    =
    \begin{bmatrix}
        \beta \\
        \alpha
    \end{bmatrix}
\end{equation}

Quantum gates that act on single qubits cannot transform a quantum register to an entangled state by themselves.
To overcome this limitation, multi-qubit quantum gates are required.
The Controlled-U or CU (where U can be any single qubit gate) is a set of quantum gates that belong to this category, as they operate on two qubits.
One qubit acts as a control qubit, while the other acts as a target qubit.
The operation is performed on the target qubit according to the state of the control qubit.
A CU gate is mathematically defined by the unitary matrix presented in Equation~\ref{eq7}.
An example of quantum gate that belongs to the CU class of gates is CNOT which operates on two qubits (Equation~\ref{eq8}).

\begin{equation}\label{eq7}
    CU = 
  \begin{blockarray}{*{4}{c} l}
    \begin{block}{*{4}{>{$\footnotesize}c<{$}} l}
      $\ket{00}$ & $\ket{01}$ & $\ket{10}$ & $\ket{11}$ \\
    \end{block}
    \begin{block}{[*{4}{c}]>{$\footnotesize}l<{$}}
      1 & 0 & 0 & 0 & $\ket{00}$ \\
      0 & 1 & 0 & 0 & $\ket{01}$ \\
      0 & 0 & U_a & U_b & $\ket{10}$ \\
      0 & 0 & U_c & U_d & $\ket{11}$ \\
    \end{block}
  \end{blockarray}
\end{equation}

\begin{equation}\label{eq8}
    CNOT = 
  \begin{blockarray}{*{4}{c} l}
    \begin{block}{*{4}{>{$\footnotesize}c<{$}} l}
      $\ket{00}$ & $\ket{01}$ & $\ket{10}$ & $\ket{11}$ \\
    \end{block}
    \begin{block}{[*{4}{c}]>{$\footnotesize}l<{$}}
      1 & 0 & 0 & 0 & $\ket{00}$ \\
      0 & 1 & 0 & 0 & $\ket{01}$ \\
      0 & 0 & 0 & 1 & $\ket{10}$ \\
      0 & 0 & 1 & 0 & $\ket{11}$ \\
    \end{block}
  \end{blockarray}
\end{equation}

\subsection{Quantum Circuits}
\label{sec:quantum_cirquits}
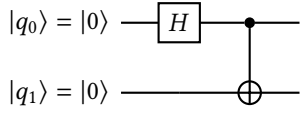
\begin{figure}[htbp!]
    \centering
    \begin{quantikz}
        \lstick{$\ket{q_0} = \ket{0}$} & \gate{H} & \ctrl{1} & \qw \\
        \lstick{$\ket{q_1} = \ket{0}$} & \qw & \targ{} & \qw  \\
    \end{quantikz}
    \caption{The Bell State quantum circuit.}
    \label{fig:bellstate}
\end{figure}

Similar to digital circuits, quantum circuits are composed of quantum gates.
In essence, quantum circuits define how quantum gates can be organized in order to be applied to the qubits of a quantum register~\cite{gruska1999, imperial_notes2015}.
Figure~\ref{fig:bellstate} presents a simplistic quantum circuit that describes a Bell state between two qubits.
The horizontal lines represent individual qubits on which quantum gates can be applied~\cite{zulehner2018}.

% In this example, we use the unitary matrix, which is a very important formula used in quantum mechanics because is preserves norms, and thus, probability amplitudes.
% % However, other simulation backends (see Section~\ref{sec:quantum_simulation_backends}) can be correspondingly applied for simulating this quantum circuit.
% Equation~\ref{eq9} presents the mathematical model of this quantum circuit after the application of the unitary matrix.
% The quantum gates that operate on individual qubits can be combined in a unitary matrix that represents the whole quantum circuit.
% This is achieved by an operation called the \textit{Kronecker product}, as expressed by symbol $\otimes$ in Equation~\ref{eq9}.
% The mathematical expression of the quantum state for this circuit is described in Appendix~\ref{sec::appendix-state-of-circuit}.

% \begin{equation}\label{eq9}
%     \ket{\psi} = \ket{0} \otimes \ket{0} = 
%     \begin{bmatrix}
%         1 \\
%         0
%     \end{bmatrix}
%     \otimes
%     \begin{bmatrix}
%         1 \\
%         0
%     \end{bmatrix}
%     =
%     \begin{bmatrix}
%         1 \cdot 
%         \begin{bmatrix}
%             1 \\
%             0
%         \end{bmatrix} \\
%         0 \cdot 
%         \begin{bmatrix}
%             1 \\
%             0
%         \end{bmatrix} 
%     \end{bmatrix}
%     =
%     \begin{bmatrix}
%         1 \\
%         0 \\
%         0 \\
%         0
%     \end{bmatrix}
%     =
%     \begin{bmatrix}
%         \alpha_{00} \\
%         \alpha_{01} \\
%         \alpha_{10} \\
%         \alpha_{11}
%     \end{bmatrix}
% \end{equation}

To mathematically perform an evaluation of the Bell state circuit (Figure~\ref{fig:bellstate}), the following stages are executed:
\begin{itemize}
    \item At first, the initial quantum state $\ket{\psi}$ is defined as $\ket{00}$, as both qubits are in the state $\ket{0}$ (Equation~\ref{eq9}).
    \item The evaluation is divided into two steps ($S_1$, $S_2$).
    \item In the first step ($S_1$ - Equation~\ref{eq9s1}), the Hadamard gate is applied to qubit 0; and the identity matrix is applied to qubit 1 (no operation on qubit 1 in this step). The Kronecker product ($\otimes$) is used to create step unitary matrix out of the individual gates.
    \item In the second step ($S_2$ - Equation~\ref{eq10}), the CNOT gate is applied to both qubits.
    \item Finally, all unitary step matrices are applied to the initial quantum state ($\ket{\psi}$) by using the standard matrix multiplication operation, as shown in Equation~\ref{eq11}.
\end{itemize}

\begin{equation}\label{eq9}
    \ket{\psi} = \ket{0} \otimes \ket{0} = 
    \begin{bmatrix}
        1 \\
        0
    \end{bmatrix}
    \otimes
    \begin{bmatrix}
        1 \\
        0
    \end{bmatrix}
    =
    \begin{bmatrix}
        1 \cdot 
        \begin{bmatrix}
            1 \\
            0
        \end{bmatrix} \\
        0 \cdot 
        \begin{bmatrix}
            1 \\
            0
        \end{bmatrix} 
    \end{bmatrix}
    =
    \begin{bmatrix}
        1 \\
        0 \\
        0 \\
        0
    \end{bmatrix}
    =
    \begin{bmatrix}
        \alpha_{00} \\
        \alpha_{01} \\
        \alpha_{10} \\
        \alpha_{11}
    \end{bmatrix}
\end{equation}

\begin{equation}\label{eq9s1}
% \[
    S_1 = H \otimes I = 
    \begin{bmatrix}
        \frac{1}{\sqrt{2}} & \frac{1}{\sqrt{2}} \\
        \frac{1}{\sqrt{2}} & \frac{-1}{\sqrt{2}}
    \end{bmatrix}
    \otimes
    \begin{bmatrix}
        1 & 0 \\
        0 & 1
    \end{bmatrix}
    =
    \begin{bmatrix}
        \frac{1}{\sqrt{2}} \cdot 
        \begin{bmatrix}
            1 & 0 \\
            0 & 1
        \end{bmatrix}
        & \frac{1}{\sqrt{2}} \cdot 
        \begin{bmatrix}
            1 & 0 \\
            0 & 1
        \end{bmatrix} \\
        \frac{1}{\sqrt{2}} \cdot 
        \begin{bmatrix}
            1 & 0 \\
            0 & 1
        \end{bmatrix}
        & \frac{-1}{\sqrt{2}} \cdot 
        \begin{bmatrix}
            1 & 0 \\
            0 & 1
        \end{bmatrix} 
    \end{bmatrix}
    =
    \begin{bmatrix}
        \frac{1}{\sqrt{2}} & 0 & \frac{1}{\sqrt{2}} & 0 \\
        0 & \frac{1}{\sqrt{2}} & 0 & \frac{1}{\sqrt{2}} \\
        \frac{1}{\sqrt{2}} & 0 & \frac{-1}{\sqrt{2}} & 0 \\
        0 & \frac{1}{\sqrt{2}} & 0 & \frac{-1}{\sqrt{2}} 
    \end{bmatrix} 
% \]
\end{equation}

\begin{equation}\label{eq10}
    S_2 = CNOT = 
    \begin{bmatrix}
        1 & 0 & 0 & 0 \\
        0 & 1 & 0 & 0 \\
        0 & 0 & 0 & 1 \\
        0 & 0 & 1 & 0 
    \end{bmatrix}
\end{equation}

\begin{equation}\label{eq11}
    \ket{\psi\prime} = S_2 \cdot S_1 \cdot \ket{\psi} = 
    \begin{bmatrix}
        1 & 0 & 0 & 0 \\
        0 & 1 & 0 & 0 \\
        0 & 0 & 0 & 1 \\
        0 & 0 & 1 & 0 
    \end{bmatrix} 
    \cdot
    \begin{bmatrix}
        \frac{1}{\sqrt{2}} & 0 & \frac{1}{\sqrt{2}} & 0 \\
        0 & \frac{1}{\sqrt{2}} & 0 & \frac{1}{\sqrt{2}} \\
        \frac{1}{\sqrt{2}} & 0 & \frac{-1}{\sqrt{2}} & 0 \\
        0 & \frac{1}{\sqrt{2}} & 0 & \frac{-1}{\sqrt{2}} 
    \end{bmatrix}
    \cdot
    \begin{bmatrix}
        1 \\
        0 \\
        0 \\
        0
    \end{bmatrix}
    =
    \begin{bmatrix}
        \frac{1}{\sqrt{2}} \\
        0 \\
        0 \\
        \frac{1}{\sqrt{2}}
    \end{bmatrix}
\end{equation}

The result is the expected Bell state: $\ket{\psi\prime} = \frac{1}{\sqrt{2}}\ket{00} + \frac{1}{\sqrt{2}}\ket{11}$.
In mathematical terms, the technique used above requires the construction of a complex $2^n \times 2^n$ matrix for each step of the quantum circuit.
Additionally, the full quantum state is represented by a complex vector of size $2^n$, where $n$ is the number of qubits in the circuit.

\subsection{Quantum Programs}
\label{sec:quantum_programs}
Programs to be executed on a quantum computer or simulated using a quantum simulator can be directly represented by a quantum circuit, which is effectively a set of steps describing the application of quantum gates on qubits within the quantum register. 
The following paragraphs describe three quantum circuits, which will be later used for the evaluation of TornadoQSim in Section~\ref{section::evaluation}.

\subsubsection{Fully Entangled Circuit}
\begin{figure}[htbp!]
    \centering
    \resizebox{0.4\columnwidth}{!}{%
    \begin{quantikz}
        \lstick{$\ket{q_0} = \ket{0}$} & \gate{H} & \ctrl{3} & \ctrl{2} & \ctrl{1} & \qw \\
        \lstick{$\ket{q_1} = \ket{0}$} & \qw & \qw & \qw & \targ{} & \qw \\
        \lstick{$\ket{q_2} = \ket{0}$} & \qw & \qw & \targ{} & \qw & \qw \\
        \lstick{$\ket{q_3} = \ket{0}$} & \qw & \targ{} & \qw & \qw & \qw \\
    \end{quantikz}
    }
    \caption{A Fully Entangled quantum circuit (4 qubits).}
    \label{fig:entanglcircuit}
\end{figure}
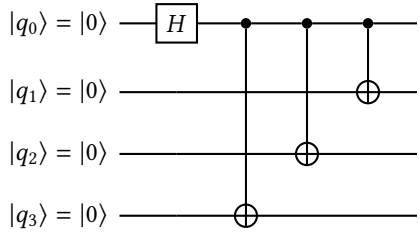

The Fully Entangled Circuit is a quantum circuit that showcases the quantum mechanical property of entanglement. 
Figure~\ref{fig:entanglcircuit} shows a graphical representation of a fully entangled quantum circuit with four qubits. 
This circuit uses a quantum register that contains four qubits, and it applies the Hadamard gate to the first qubit. 
Then, the next three steps include the application of the CNOT gate to qubits 3, 2, and 1, respectively.
Note that all CNOT gates are conditioned on the first qubit.

\subsubsection{The Deutsch-Jozsa Quantum Algorithm}
The Deutsch-Jozsa Quantum algorithm \cite{51_deutsch} is a $n$-bit generalization of the Deutsch algorithm \cite{53_qiskitdeutsch}. 
It is one of the algorithms to showcase the quantum advantage when compared to the best known classical algorithm for solving the same problem.
% The Deutsch-Jozsa algorithm outperforms the best known classical algorithm for this problem, and therefore, serves as a good demonstration of the potential speedup that quantum computers promise \cite{52_qcforcs}.
Given a boolean function $f$ (Equation~\ref{eqdj}) that takes a binary input of $n$ bits, the algorithm outputs a single bit output that determines whether the function is \textit{balanced} or \textit{constant}.
The function $f$ is \textit{balanced} if the single-bit output is `0' for exactly half of the inputs, and `1' for the other half of the inputs.
Alternatively, the function is \textit{constant} if the single-bit output is `0' (or `1') for all of the input combinations.
% Table~\ref{table:functionout} presents an example of a truth table for both the balanced and constant functions using three bits as input.
\begin{equation}\label{eqdj}
f(x_n, x_{n-1},\dots, x_0) \mapsto \{0,1\}, x_n \in \{0,1\}
\end{equation}

% \begin{table}[htbp!]
%     \centering
%     \begin{tabular}{c c c}
%     \toprule
%     Input (x\_2x\_1x\_0) & Balanced & Constant  \\  
%     \midrule
%     000 & 0 & 0 \\
%     \midrule
%     001 & 1 & 0 \\ 
%     \midrule
%     010 & 1 & 0 \\
%     \midrule
%     011 & 0 & 0 \\
%     \midrule
%     100 & 1 & 0 \\
%     \midrule
%     101 & 0 & 0 \\ 
%     \midrule
%     110 & 0 & 0 \\
%     \midrule
%     111 & 1 & 0 \\
%     \bottomrule 
%     \end{tabular}
%     \caption{Truth table of balanced and constant functions.}
%     \label{table:functionout}
% \end{table}

Typically, the problem of finding whether the type of a function is \textit{balanced} or \textit{constant} can be solved by trying different input combinations.
The best scenario is when the output of the second trial is different from the output of the first trial.
In that case, it is clear that the function cannot be constant and the algorithm can be stopped.
However, more than half of the input combinations must be tried in the worst case, to rule out the possibility of the function being balanced.
In general terms, this is $2^{n-1} + 1$ trials, where $n$ is the number of input bits \cite{53_qiskitdeutsch}.

\begin{figure}[htbp!]
    \centering
    \resizebox{0.4\columnwidth}{!}{%
    \begin{quantikz}
        \lstick{$\ket{q_0} = \ket{0}$} & \gate{H} &  \gate[wires=4][2cm]{U_f} 
            \gateinput[3]{$x$}
            \gateoutput[wires=3]{$x$} & \gate{H} & \meter{} & \qw \\
        \lstick{$\ket{q_1} = \ket{0}$} & \gate{H} & & \gate{H} & \meter{} & \qw  \\
        \lstick{$\ket{q_2} = \ket{0}$} & \gate{H} & & \gate{H} & \meter{} & \qw  \\
        \lstick{$\ket{q_3} = \ket{1}$} & \gate{H} & \gateinput{$y$}\gateoutput{$y\oplus f(x)$} & \qw & \qw & \qw  \\
    \end{quantikz}
    }
    \caption{A Deutsch-Jozsa Quantum circuit (4 qubits, 3 function inputs).}
    \label{fig:deutschcircuit}
\end{figure}
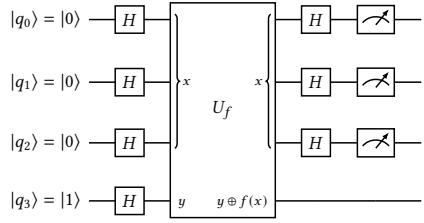

The quantum version of this algorithm only requires a single evaluation to determine the type of the function \cite{53_qiskitdeutsch}.
The quantum circuit that implements the algorithm for a function of three inputs is shown in Figure~\ref{fig:deutschcircuit}. The main component of the quantum circuit is the quantum oracle $U_f$, which is a provided piece of quantum circuit that implements the function to be decided. However, the implementation details cannot be viewed, and it is considered as a black box \cite{54_oracle}.
Once the quantum circuit is evaluated, all qubits from 0 to n will collapse to the state 0 or state 1 if the function is balanced or constant, respectively \cite{53_qiskitdeutsch}.

\subsubsection{Quantum Fourier Transform}
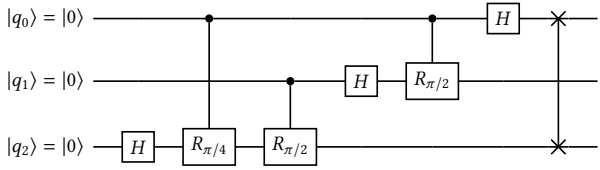
\begin{figure}[htbp!]
    \centering
    \resizebox{0.6\columnwidth}{!}{%
    \begin{quantikz}
        \lstick{$\ket{q_0} = \ket{0}$} & \qw & \ctrl{2} & \qw & \qw & \ctrl{1} & \gate{H} & \swap{2} & \qw & \\
        \lstick{$\ket{q_1} = \ket{0}$} & \qw & \qw & \ctrl{1} & \gate{H} & \gate{R_{\pi/2}} & \qw & \qw & \qw & \\
        \lstick{$\ket{q_2} = \ket{0}$} & \gate{H} & \gate{R_{\pi/4}} & \gate{R_{\pi/2}} & \qw & \qw & \qw & \swap{0} & \qw & \\
    \end{quantikz}
    }
    \caption{A QFT Quantum circuit (4 qubits).}
    \label{fig:qftcircuit}
\end{figure}

The Quantum Fourier transform (QFT) is the basic building block of some of the more complex quantum algorithms, such as the Shor's algorithm that demonstrates an efficient solution of integer-factorization on a quantum computer \cite{53_qiskitqft}.
The QFT algorithm transforms the quantum state $\ket{x}$ from the computational basis to the Fourier basis \cite{53_qiskitqft}.
The quantum state in Fourier basis is often mathematically expressed as shown in Equation~\ref{eqqft}.
Figure~\ref{fig:qftcircuit} illustrates a circuit of three qubits, where QFT is applied.
\begin{equation}\label{eqqft}
    QFT\ket{x} = \ket{\tilde{x}}
\end{equation}

\subsection{Simulation Backends}
\label{sec:quantum_simulation_backends}
The most straightforward approach for simulating a quantum circuit on a conventional computer has been the direct application of the mathematical model (i.e.,\ using Kronecker product and matrix multiplication), as mentioned in the previous section.
However, this method requires the construction of a unitary matrix for each step of the quantum circuit, thereby increasing the memory space exponentially with the number of qubits~\cite{iqs2020}.
For instance, the minimum space for simulating a quantum circuit with 20 qubits requires to store $2^{20}$ probability amplitudes of the state vector, and a $2^{20} \times 2^{20}$ unitary matrix of at least one step in the quantum circuit.
If a 64-bit representation is considered for each complex number, this will result in a required capacity of around 8.8 TB for a successful simulation.
% Additionally, an exponential number of arithmetic operations, $2^n(2\cdot2^n - 1)$, is required to apply a step of the quantum circuit \citep{2_QC_Gruska}.

Other simulation techniques have been employed as simulation backends due to their efficacy in memory space~\cite{iqs2020}.
For instance, in case of circuits that have separable states (no two-qubit states), all the quantum gates can be separately applied without the need to construct a unitary matrix for each step~\cite{hidary2019}.
% These circuits are used by \textit{unitary matrix simulators}~\cite{qiskit_docs2021}, but they are not widely used because they do not exploit the quantum entanglement principle.
% \ta{The wording here (as was in my report) is not ideal it implies things that are incorrect}
Another optimization upon the unitary matrix technique is to overcome the construction of a unitary matrix, but rather apply the quantum gates directly on the full state vector~\cite{hidary2019, qdk_docs2021}.
This paper gives more emphasis on the unitary matrix simulation, thus other simulation techniques are discussed in Appendix~\ref{sec::appendix-fsv-tensor}.

%Same notation is used to describe unitary matrices that represent quantum gates.
% Moreover, the Kronecker product and matrix multiplication operations are defined to be applied on the decision diagram, allowing efficient simulation of universal quantum circuits~\cite{zulehner2018, viamontes2003}.
% \ts{The last sentence is not very clear. Ales, can you please explain?}
% \ta{Yes - during the meeting}

\section{Background in Hardware Acceleration of Managed Programming Languages}
\label{sec:hardware_acceleration}
The state-of-the-art simulators employ hardware acceleration as a means to reduce the simulation time of the quantum circuits~\cite{mccaskey2018, qflex2019}.
To achieve this, the vast majority of the available simulation frameworks employs GPUs~\cite{qdk_docs2021, quest2019} and FPGAs~\cite{lee2016, pilch2019}.
Both hardware devices are widely used as co-processors to the main CPU in order to offload computations that can be performed in parallel.
GPUs are suitable for applications that offer loop parallelism, such as computer vision ~\cite{10.1145/1101149.1101334, 10.1145/3050748.3050764} and deep learning ~\cite{227623, 9355309, 10.1145/3426182.3426188}.
On the other hand, FPGAs are best for applications that can exhibit pipeline parallelism, due to the combination of various on-device resources (i.e.,\ blocks of memory, registers, logic slices) that can be dynamically reconfigured to compose custom hardware blocks ~\cite{10.1145/508352.508353}.
For instance, financial technology ~\cite{vitis_fintech_2021} is a domain that can exploit FPGAs for accelerating math operations (e.g., sine and cosine),
as FPGAs can execute these operations in few clock cycles.

\subsection{Transparent Hardware Acceleration with TornadoVM}
\label{sec:hardware_acceleration_tornadovm}
Although heterogeneous hardware accelerators offer high performance, they come at a high cost with regards to programming.
To facilitate programming, several programming models have been developed as a standardized way to program these niche hardware types, such as OpenCL~\cite{opencl:2022}, CUDA~\cite{nvidia-cuda}, OneAPI~\cite{opencl:2022}.
However, these programming models are supported by unmanaged compiled languages (i.e.,\ C/C++) and are designed for programmers who have hardware knowledge in order to exploit more features (e.g.,\ custom memory access). 

In recent years, numerous programming frameworks (e.g.,\ TornadoVM) have emerged to aid software programmers making their applications capable of harnessing heterogeneous accelerators.
TornadoVM~\cite{tornado2018, tornado2019} is a state-of-the-art heterogeneous programming framework that can transparently translate Java Bytecodes to OpenCL and PTX.
Additionally, TornadoVM applies various code specialization techniques that can automatically tailor the generated code to the characteristics of the targeted hardware co-processor (CPU, GPU, FPGA)~\cite{tornado2019}.
Therefore, Java applications can be seamlessly accelerated, without requiring a programmer to code any low-level platform-specific code in CUDA or OpenCL or any other heterogeneous programming language.

TornadoVM offers the \texttt{TaskSchedule} API, a Java API that allows software programmers to model the parallel execution of their applications on heterogeneous hardware.
A \texttt{TaskSchedule} is a group of acceleratable tasks, in which each task is semantically equivalent to a Java method and an OpenCL kernel.
Listing~\ref{listing:tornadoschedule} presents an example where the \texttt{TaskSchedule} is used to offload a vector addition Java method (Listing~\ref{listing:tornadomethod}) on a hardware accelerator.
At the beginning, a new object of \texttt{TaskSchedule} is created in Line 2, while Lines 3 and 5 define which data are used by the method as inputs and outputs, respectively.
Subsequently, Line 4 defines a task which accepts as input a String identifier (e.g.,\ ``t0''), a reference to the acceleratable method (e.g.,\ this::vadd) along with the arguments of this method (e.g.,\ a, b, c).
Finally, Line 5 invokes the compilation of the method to OpenCL/PTX and the execution on the underlying hardware device.

\begin{lstlisting}[caption={Example of Vector Addition Using a TaskSchedule.},label={listing:tornadoschedule},captionpos=b,language=Java]
    public void compute(int[] a, int[] b, int[] c) {
      TaskSchedule s = new TaskSchedule ("s0")
        .streamIn(a, b)
        .task("t0", this::vadd, a, b, c)
        .streamOut(c)
        .execute();
    }
\end{lstlisting}

Additionally, the TornadoVM API exposes annotations that can be used in the acceleratable methods for parallel programming.
In particular, TornadoVM exposes the \texttt{@Parallel} annotation to mark loops that can be candidate for parallel execution~\cite{tornado2018}.
Listing~\ref{listing:tornadomethod} presents the annotated Java method for computing vector addition.
The code shown in Listing~\ref{listing:tornadomethod} is Java code, with the only difference that the loop uses the \texttt{@Parallel} annotation in Line 2.

\begin{lstlisting}[caption={Example of the Vector Addition Method in Java Using the TornadoVM Annotation.},label={listing:tornadomethod},captionpos=b,language=Java]
public void vadd(int[] a, int[] b, int[] c) {
    for (@Parallel int i = 0; i < c.length; i++) {
        c[i] = a[i] + b[i];
    }
}
\end{lstlisting}

% Section~\ref{sec:related_work} presents a detailed overview of the simulators that are relevant to this paper.

\section{Related Work}
\label{sec:related_work}
Quantum simulators typically provide an Application Programming Interface (API) to define and optimize quantum circuits, while allowing users to select a simulation backend to perform the actual simulation of a quantum circuit.
A backend can be a simulation model for a quantum circuit or a connector to a real quantum computer~\cite{quest2019}. 
The API that allows users to easily define quantum circuits is usually implemented in a high-level and widely used language~\cite{hidary2019},
whereas the quantum simulation backends are typically implemented in low-level languages due to the prospect of obtaining higher performance~\cite{qflex2020, projectq2017}.

This section groups the related work into two groups.  
The first class presents the state-of-the-art quantum simulators that use Python and Q\# for interfacing, while also supporting both a single backend and multiple backends.
The second class presents a subset of the quantum simulators that use Java for both the user programming interface and the applied simulation backend.
The second category is tightly related to this paper.

\subsection{State-of-the-art in Quantum Simulators}
\label{sec:sota_simulators}
Quantum simulators present a high demand in computation as the space for representing a circuit increases exponentially to the number of qubits.
To satisfy the need for fast computation in large scale, current state-of-the-art simulators usually employ High Performance Computing (HPC) techniques.
For instance, the MPI parallel programming model has been used as a HPC technique to distribute the problem across many computing nodes~\cite{qflex2019, iqs2020, ibmsim2017, quest2019, qflex2020, qhipster2016, projectq2017, projectq2018}.
Similarly, OpenMP has been employed to exploit local computation speedup by applying multi-threaded execution~\cite{ibmsim2017, quest2019, qhipster2016, projectq2017}.
Other works have mapped some calculations to GPGPU kernels (CUDA, OpenCL) ~\cite{qdk_docs2021, quest2019} or have applied vectorization and other low-level mathematical routines (BLAS, MKL, AVX)~\cite{qflex2019, ibmsim2017, qflex2020, projectq2017, projectq2018}.
Furthermore, other simulators, such as QuEST~\cite{quest2019}, can dynamically switch between the computation techniques based on the size of the quantum circuit.
With these optimizations, the current state-of-the-art simulators are able to simulate circuits of 45 or even 49 qubits on supercomputers with memory requirements in the order of petabytes~\cite{ibmsim2017, projectq2017}.
Table~\ref{table:single_backend} presents a synopsis of the state-of-the-art simulators along with their implementation details (simulation backend, optimizations, programming languages).
As shown in Table~\ref{table:single_backend}, all simulators implement one prime simulation backend and they use primarily the C/C++ programming language for the backend part of the simulator, while the majority also exposes an API in Python.

\begin{table}[htbp!]
    \centering
    \scalebox{0.9}{
    \begin{tabular}{l p{4cm} p{2.8cm} p{3.2cm}}
    \toprule
    Name & Simulation Backend & Optimization & Language  \\  
    \midrule
        --- (IBM) \cite{ibmsim2017} & 
        Tensor Network & 
        MPI \newline OpenMP \newline BLAS &
        C++ (backend) \\
    \midrule
        QuEST \cite{quest2019} & 
        Full State Vector & 
        MPI \newline OpenMP \newline CUDA &
        C (backend) \newline CUDA (backend) \newline Python (API) \\
    \midrule
        qFlex \cite{qflex2019, qflex2020} & 
        Tensor Network \newline (TAL-SH) & 
        CUDA \& cuBLAS, \newline MKL &
        C/C++ (backend) \newline Python (API) \\
    \midrule
        IQS \cite{iqs2020} & 
        Full State Vector & 
        MPI \newline OpenMP &
        C++ (backend) \newline Python (API) \\
    \midrule
        ProjectQ \cite{projectq2017, projectq2018} & 
        Full State Vector & 
        MPI \newline OpenMP \newline AVX &
        C++ (backend) \newline Python-DSL (API, Compiler) \\
    \midrule
        JKQ DDSIM \cite{zulehner2018} & 
        QMDD & 
        - &
        C++ (backend) \\
    \bottomrule 
    \end{tabular}
    }
    \caption{Summary of state-of-the-art single backend quantum simulators.}
    \label{table:single_backend}
\end{table}

Unlike the above-mentioned quantum simulators that focus on a single simulation backend, other quantum frameworks provide support for multiple backends.
Table~\ref{table:multiple_backends} presents the state-of-the-art quantum computing frameworks along with the supported backends and the programming language of the API.
The Qiskit, Cirq, and Quantum Development Kit (QDK) frameworks have been developed by IBM, Google and Microsoft, respectively.
Qiskit and Cirq expose a Python API, while QDK exposes an API in Q\# and offers interoperability with workflows from Cirq and Qiskit.
Finally, all frameworks allow users to run circuits on real quantum systems.

\begin{table}[htbp!]
    \centering
    \scalebox{0.9}{
    \begin{tabular}{l p{8cm} p{3.2cm}}
    \toprule
    Name & Simulation Backend & Language \\  
    \midrule
        Qiskit \cite{qiskit_docs2018} & 
        AerSimulator (unitary, statevector, \newline 
        density\_matrix, stabilizer, \newline
        matrix\_product\_state, etc.) \newline &
        C++ (backend) \newline Python (API) \\
    \midrule
        Cirq \cite{cirq_docs2021} &
        Built-in \textit{pure} state and \textit{mixed} state \& \newline 
        External Simulators: \newline 
        qsim (Full State Vector) \newline 
        qsimh (Full State Vector) \newline 
        qFlex (Full State Vector) \newline 
        quimb (Tensor Network) \newline &
        Python (API \& backend) \\
    \midrule
        QDK \cite{qdk_docs2021} & 
        QuantumSimulator (Full State Vector) \newline 
        SparseSimulator \newline
        ResourcesEstimator \newline
        QCTraceSimulator \newline
        ToffoliSimulator \newline 
        OpenSystemsSimulator \newline &
        Q\# (API) \\
    \bottomrule 
    \end{tabular}
    }
    \caption{Summary of state-of-the-art multi-backend quantum frameworks.}
    \label{table:multiple_backends}
\end{table}

\subsection{Quantum Simulators and Java}
In recent years, there has been a number of quantum simulators implemented using the Java programming language~\cite{quantiki2021}.
However, only few of them are still active.
%Four out of ten of these simulators seem to be still accessible. 
Qubit101~\cite{qubit101_docs2021} and jQuantum~\cite{jquantum_docs2010} provide a user interface to define quantum circuits, while LibQauntumJava (LQJ)~\cite{lqj_docs2018} and Strange~\cite{strange_docs2021} expose an API to users for defining the quantum circuit directly from Java.
All four simulators implement a single simulation backend, without applying any additional acceleration.

The most promising quantum simulator in Java is Strange~\cite{strange_docs2021}. %, which seems to be the only one under active development.
Strange has developed an architecture for enabling the user to switch between multiple simulation environments (simulation backends)~\cite{strange_docs2021}, thus it can be categorized as quantum simulation framework.
However, the drawback of Strange is the API, which forces the developer to explicitly define steps in quantum circuits, instead of handling this automatically as other frameworks do (e.g.,\ Qiskit~\cite{qiskit_docs2021} or Cirq~\cite{cirq_docs2021}).
Another disadvantage is the lack of hardware acceleration, unlike other simulators mentioned in Section~\ref{sec:sota_simulators}.
On the other hand, StrangeFX~\cite{strangefx} is being emerged as a companion project to allow a visualization and graphical definition of quantum circuits via a drag-and-drop graphical user interface.

In this paper, we aim to advance prior work on implementing quantum simulation in Java, by introducing TornadoQSim (Section~\ref{sec:framework}); an open-source framework that implements a modular architecture and employs transparent hardware acceleration.
\section{The TornadoQSim Framework}
\label{sec:framework}
TornadoQSim is an open-source framework that is developed in Java to be modular and easily expandable for accommodating various user-defined quantum simulation backends (e.g.,\ the unitary matrix simulation backend).
Section~\ref{sec:system_overview} presents the overall architecture of TornadoQSim. 
Section~\ref{sec::simulation_interfaces} describes a specification of the key software interfaces.
Section~\ref{sec::simulation_flow} outlines the execution flow for simulating an example circuit, whereas Section~\ref{sec:modularity} shows how the modular architecture of TornadoQSim can be expanded to support various simulation backends.
Finally, Section~\ref{sec:system_hw_accel} discusses how TornadoQSim exploits transparent hardware acceleration of its backends on heterogeneous architectures.

\subsection{System Overview}
\label{sec:system_overview}
\begin{figure}[htbp!] 
    \centering    
    \includegraphics[width=0.7\textwidth]{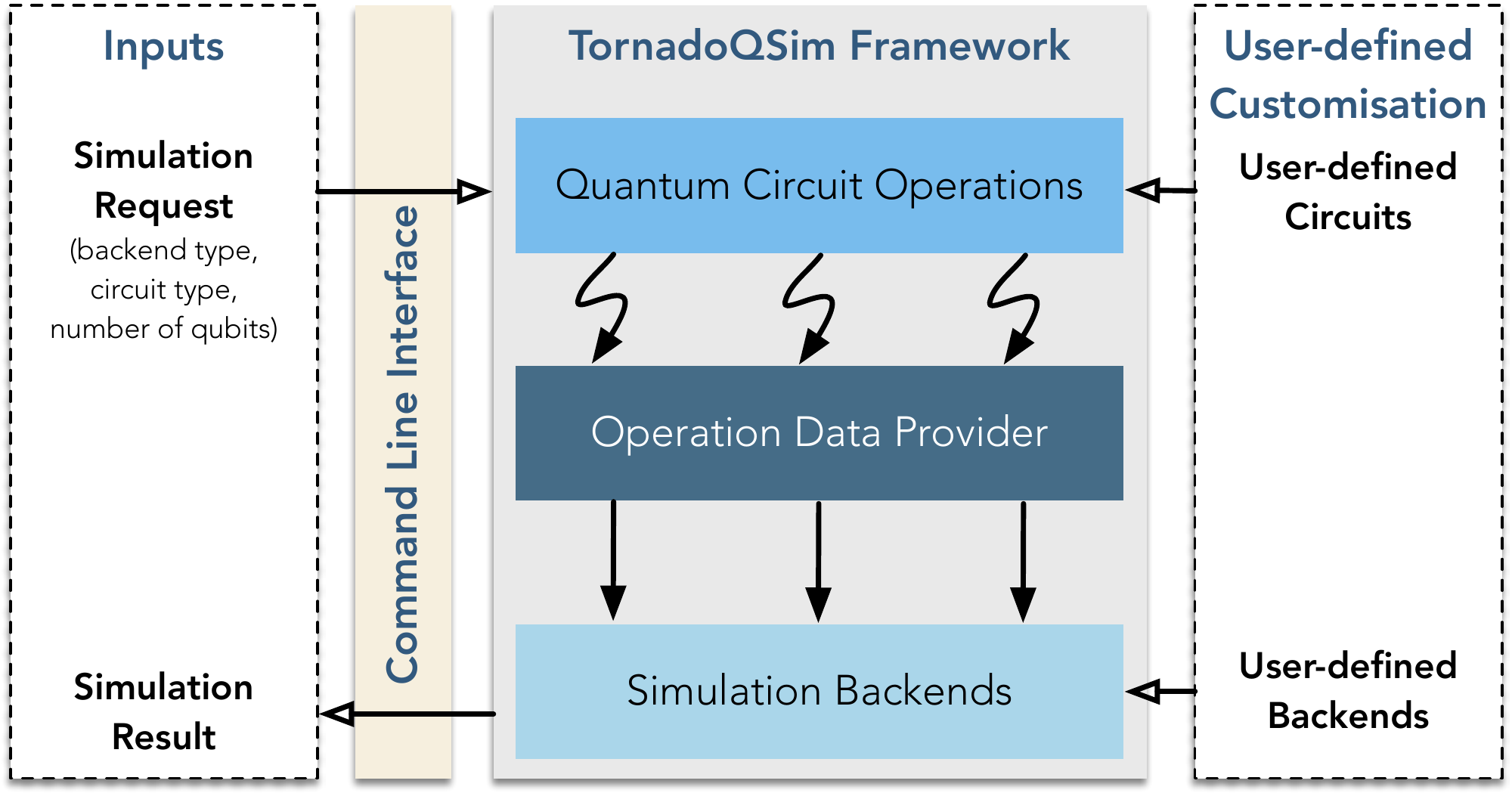}
    \caption{The TornadoQSim architecture.}
    \label{fig:architecture}
\end{figure}

The TornadoQSim architecture is designed to decouple the quantum circuits from the simulation backends.
Figure~\ref{fig:architecture} presents the overall TornadoQSim architecture that comprises three core modules, the \textit{Quantum Circuit Operations}, the \textit{Operation Data Provider}, and the \textit{Simulation Backends}.
Additionally, TornadoQSim provides a command line interface that allows users to configure the simulation by submitting a request to simulate a circuit of user-defined qubits with a user-defined simulation backend.
Finally, TornadoQSim is architected to facilitate users to easily expand the current set of circuits or backends for simulating their quantum circuits. 

\subsection{TornadoQSim Specification of Software Interfaces in Java}
\label{sec::simulation_interfaces}
The model of a quantum circuit in TornadoQSim is designed to be decoupled from the simulation backend.
This is not always the case for other implementations.
For instance, Strange~\cite{strange_docs2021} stores the complex unitary matrix of each quantum gate directly in the structure of the quantum circuit.
If a new simulation backend with a different representation of quantum gates would be later added to the framework (e.g.,\ tensor network backend), the circuit structure would need to be refactored.
TornadoQSim overcomes this problem with an architectural decision to model the circuit structure independently of any quantum gate data.
Instead, the data for each type of a quantum gate is supplied to a simulation backend by an \textit{Operation Data Provider} (Figure~\ref{fig:architecture}).
% The role of the provider will be described later in Section~\ref{sec::operation_provider}.
% At this stage, it can be considered as an intermediate data structure that stores the data of a quantum gate.
% This module serves as an intermediate layer that allows the simulation backends to request data of different operations (Figure~\ref{fig:architecture}).
% This is a key design choice of the TornadoQSim architecture as a means to enable any quantum circuit to be modeled independently of any quantum gate.
% Unlike TornadoQSim, other frameworks store the backend information of each quantum gate along with the structure of the quantum circuit~\cite{strange_docs2021}.

The following paragraphs present the TornadoQSim interfaces that are implemented by its three core modules (Figure~\ref{fig:architecture}), 
\textit{Quantum Circuit Operations} (Section~\ref{sec::circuit_model}), \textit{Operation Data Provider} (Section~\ref{sec::circuit_operations}), and \textit{Simulation Backends} (Section~\ref{sec::simulator_backends}).

\subsubsection{The Circuit Model Interface}
\label{sec::circuit_model}
\begin{figure}[htbp!] 
    \centering    
    \includegraphics[width=0.6\textwidth]{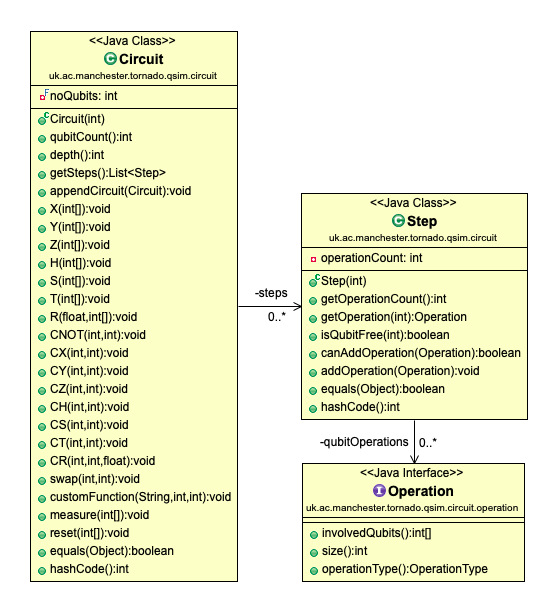}
    \caption{UML diagram of the circuit model representation.}
    \label{fig:umlcircuit}
\end{figure}

Figure~\ref{fig:umlcircuit} shows the unified modeling language (UML) diagram of the TornadoQSim circuit model. 
A user can create a new circuit of $n$ qubits and apply the basic quantum operations on the circuit by specifying qubits on which the operation should be performed. 
TornadoQsim then dynamically creates new steps when quantum operations are added to the circuit. 
The only way to create an operation and add it to the quantum circuit is by using the predefined methods of the circuit model. 
This enables to efficiently monitor the preconditions of the quantum operations as well as to hide the process of creating a new operation from a user, and thus, offering a more user-friendly API similar to the state-of-the-art frameworks (e.g.,\ Qiskit~\cite{qiskit_docs2021}).

\subsubsection{The Quantum Circuit Operation Interface}
\label{sec::circuit_operations}
There are multiple types of quantum operations supported by TornadoQSim.
The first type is a single qubit gate that applies a unitary operation on the specified qubit.
Supported types of single qubit gates in TornadoQSim are X, Y, Z, H, S, T and an arbitrary phase gate R (Section~\ref{sec:quantum_gates}), where the phase rotation is defined by a user.
Another type of an operation is a controlled gate, or equivalently two-qubit gate, where a user specifies the target qubit and the control qubit.
As the implemented set of single qubit and two-qubit gates is universal, TornadoQSim can be used to represent any possible quantum circuit.

\begin{figure}[htbp!] 
    \centering    
    \includegraphics[width=0.6\textwidth]{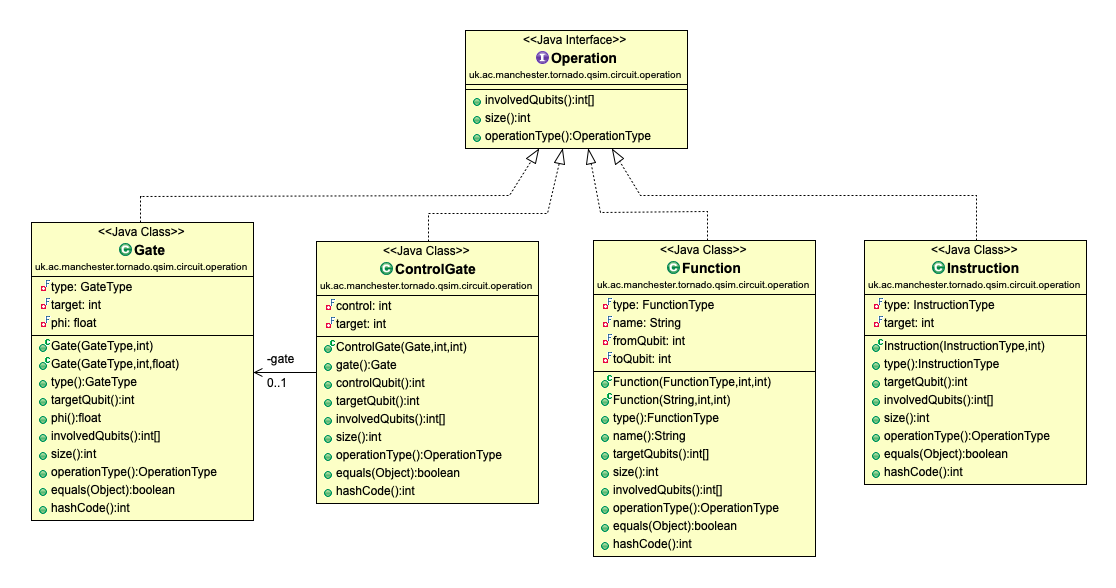}
    \caption{UML diagram of the quantum operation structure.}
    \label{fig:umloperation}
\end{figure}

Custom unitary operations can also be defined in TornadoQSim.
First, the data of a custom function needs to be registered with the \textit{Operation Data Provider}.
Then, the custom function can be applied to a range of qubits in the quantum circuit.
The function is identified by a unique name and when inserted into the quantum circuit, TornadoQSim checks if unitary data of a correct format and size were registered with the data provider prior to the insertion.
The last operation type is a single qubit instruction, typically used to control the quantum hardware rather than to manipulate qubits.
For instance, this can be a measurement instruction or a reset instruction to restore the qubit to an initial state.
The UML diagram capturing the different implementations of a general operation interface is shown in Figure~\ref{fig:umloperation}.

\subsubsection{The Simulator Interface \& Simulation Backends}
\label{sec::simulator_backends}
The final key interface of TornadoQSim is the \textit{Simulator} interface that establishes an easy integration of user-defined backends within the whole simulation framework.
That interface declares two methods, named \textit{simulateFullState} and \textit{simulateAndCollapse}.
The former method invokes a simulation process of the supplied circuit and returns the full state of the quantum system.
While, the latter method simulates the supplied circuit and it returns only the collapsed state.
In this case, the state of the circuit is collapsed (i.e.,\ measurement is performed) based on the probability that emerges from the state vector.

\begin{figure}[htbp!] 
    \centering    
    \includegraphics[width=0.6\textwidth]{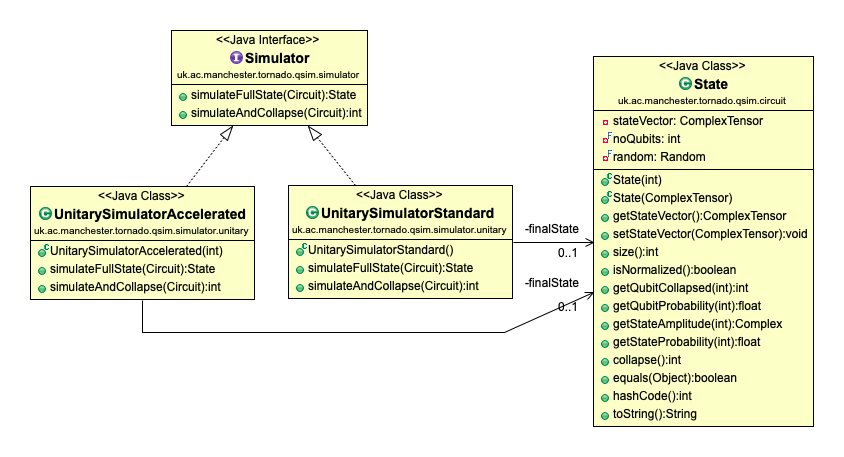}
    \caption{UML diagram of the simulation backend representation.}
    \label{fig:umlsimulator}
\end{figure}

Figure~\ref{fig:umlsimulator} presents the UML diagram of the implemented backends. 
Currently, TornadoQSim provides two simulation backends that implement the \textit{Simulator} interface with the unitary matrix simulation type.
The only difference between the two backends is that the \textit{UnitarySimulatorAccelerated} backend employs hardware acceleration through TornadoVM,
while \textit{UnitarySimulatorStandard} refers to the original Java implementation of the unitary matrix technique.

% Additionally, the model of the quantum state allows to query individual probability amplitudes as well as probabilities of the system collapsing to a specific state. 
% It should be noted that the ordering of the qubits in the state is different to most of the textbooks and examples above. 
% Usually, $\ket{100}$ represents a quantum state where the first qubit (qubit 0) is set to one, while qubits 1 and 2 are set to zero. 
% However, this does not correspond to the representation of classical bits, where bit 0 is typically the rightmost bit. 
% To make the notation consistent, the ordering of qubits for TornadoQSim is the same as for classical bits; $\ket{100}$ represents a quantum state where qubit 0 and 1 are set to zero and qubit 2 to one. 
% The same notation is also used by Qiskit~\cite{qiskit_docs2021}.

\begin{figure}[htbp!] 
    \centering    
    \includegraphics[width=0.8\textwidth]{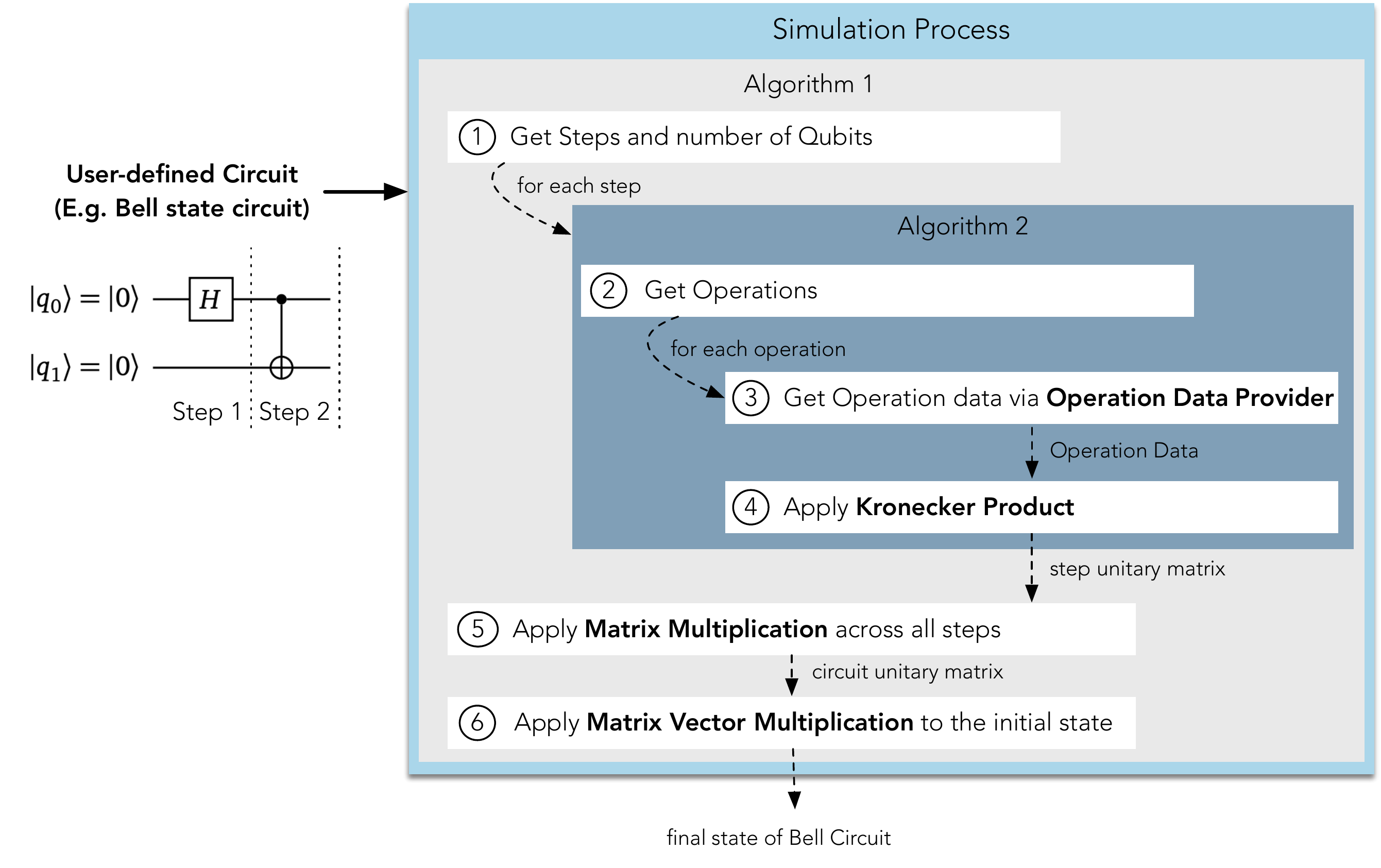}
    \caption{Overview of the simulation process using unitary matrix simulation in TornadoQSim.}
    \label{fig:simulation-flow}
\end{figure}

\subsection{Execution Flow of Simulation}
\label{sec::simulation_flow}
% Currently, TornadoQSim provides two simulation backends that implement the \textit{Simulator} interface with the unitary matrix simulation type.
% The only difference between the two backends is that the \textit{UnitarySimulatorAccelerated} backend employs hardware acceleration through TornadoVM,
% while \textit{UnitarySimulatorStandard} refers to the original Java implementation of the unitary matrix technique.
The execution of a simulation in TornadoQSim is an iterative process that goes over the steps contained in the simulated circuit.
Figure~\ref{fig:simulation-flow} presents an abstracted view of the simulation flow that is implemented in the unitary matrix simulation backend of TornadoQsim.
The simulation flow is composed of six main stages, as follows.
Stage \circled{1} retrieves the steps and the number of qubits of the simulated circuit. 
For example, a Bell state quantum circuit contains two steps, as discussed earlier in Section~\ref{sec:quantum_cirquits}.
Stage \circled{2} is invoked to obtain the operations that will be applied in each step.
Then stage \circled{3} is responsible to get all operation data that corresponds to each operation.
Following our example, the unitary matrix of the Hadamard gate is returned for the application of Hadamard gate to the qubit 0 (Step 1).
Whereas, the unitary matrix of the CNOT gate is returned for the application of the CNOT gate to both qubits (Step 2).
Stage \circled{4} applies the Kronecker product over the operation data and produces the unitary matrix for a particular step.
Subsequently, stage \circled{5} applies a matrix multiplication across all unitary matrices that have been returned from the previous stage.
Finally, stage \circled{6} performs a matrix vector multiplication between the initial state and the result of stage \circled{5}, which is the circuit unitary matrix.
The outcome of this mathematical operation is the final Bell state.
Note that other simulation backends may modify the mathematical calculations (e.g.,\ matrix multiplication, Kronecker product, etc.) in stages \circled{4}, \circled{5}, and \circled{6} to implement different simulation techniques.

As shown in Figure~\ref{fig:simulation-flow}, the six stages can be also grouped into two algorithmic parts which are expressed in Algorithms ~\ref{algo:unitary} and ~\ref{algo:step}.
Algorithm~\ref{algo:unitary} presents a generic description of simulating a quantum circuit by applying the unitary matrix technique, which 
initializes a state with the identity matrix and then it performs an iterative call to the $get\_step\_unitary(step)$ function for each circuit step.
The $get\_step\_unitary(step)$ function implements the construction of a step unitary matrix ($A$), as presented in Algorithm~\ref{algo:step}.
The key additional components for the architecture of this simulation backend are \textit{UnitaryDataProvider} and \textit{UnitaryOperand}.
The former component is responsible for preparing all unitary matrices in the correct order based on the supplied quantum circuit step.
In addition, unitary matrices for two-qubit gates that are applied on non-adjacent qubits have to be dynamically created.
The \textit{UnitaryDataProvider} takes care of this process and supplies a list of ready to use unitary matrices, which are then directly used to construct the final step unitary matrix by using the Kronecker product.
The latter component, \textit{UnitaryOperand}, encapsulates the actual mathematical calculations that need to be performed in the simulation process; namely, matrix multiplication, matrix vector multiplication and Kronecker product.
Moreover, it implements a method for the dynamic construction of the two-qubit gate unitary matrices.
All of these methods in the \textit{UnitaryOperand} model are good candidates to be accelerated on heterogeneous co-processors, as parallelism can be exploited for those operations, thus providing potential performance speedup. 

With the support of these two additional components, Algorithm~\ref{algo:unitary} is then directly implemented in both simulation backends (\textit{UnitarySimulatorStandard} and \textit{UnitarySimulatorAccelerated}).
The primary difference of the \textit{UnitarySimulatorAccelerated} model is that the accelerated methods are invoked using a TaskSchedule, which is a part of TornadoVM (Section~\ref{sec:hardware_acceleration}).

    \begin{algorithm}[H]
    \SetAlgoLined
        
        $result \leftarrow I$\;
        \ForEach{$step$ in $circuit\_steps$}{
            $A \leftarrow get\_step\_unitary(step)$\; 
            $result \leftarrow A \cdot result$\;
        }
        \KwRet{$result \cdot initial\_state\_vector$}
    \caption{Unitary matrix simulation technique.}
    \label{algo:unitary}
    \end{algorithm}
        
    \begin{algorithm}[H]
    \SetAlgoLined
        
        $operation\_list \leftarrow get\_operations\_for\_step(step)$\; 
        $result \leftarrow operation\_list\left[0\right]$\;
        \ForEach{$operation$ in $operation\_list\left[1 \rightarrow \left(n-1\right)\right]$}{
            $result \leftarrow result \otimes operation$\;
        }
        \KwRet{$result$}
    \caption{Construction of a step unitary matrix.}
    \label{algo:step}
    \end{algorithm}

Additionally, Listing~\ref{listing:belltornado} presents how to define and simulate a simple Bell quantum circuit (Figure~\ref{fig:simulation-flow}) by using TornadoQSim. 
Line 1 defines the number of qubits that exist in the Bell state circuit.
Line 3 creates a new Circuit object that contains the number of qubits.
Lines 4-5 correspond to the two steps of the Bell state circuit. The first step applies the Hadamard gate to qubit 0, while the second step applies the CNOT gate to both qubits.
Line 7 creates the simulator object that will be used for the simulation of the circuit.
Finally, line 8 performs the actual simulation.

\begin{lstlisting}[caption={Bell state quantum circuit expressed by using TornadoQSim.},label={listing:belltornado},captionpos=b,language=Java]
int noQubits = 2;

Circuit circuit = new Circuit(noQubits);
circuit.H(0);
circuit.CNOT(0, 1);

Simulator simulator = new UnitarySimulatorStandard();
State fullState = simulator.simulateFullState(circuit);
\end{lstlisting}

For completion, we present Listings~\ref{listing:bellstrange} and ~\ref{listing:bellqiskit} that present the definition of the functionally equivalent programs using Strange and Qiskit, respectively.

\begin{lstlisting}[caption={Bell state quantum circuit expressed by using Strange.},label={listing:bellstrange},captionpos=b,language=Java]
int noQubits = 2;
Program program = new Program(noQubits);

Step step1 = new Step();
step1.addGate(new Hadamard(0));
program.addStep(step1);

Step step2 = new Step();
step2.addGate(new Cnot(0, 1));
program.addStep(step2);

SimpleQuantumExecutionEnvironment environment 
  = new SimpleQuantumExecutionEnvironment();
Result result = environment.runProgram(program);
\end{lstlisting}

\begin{lstlisting}[caption={Bell state quantum circuit expressed by using Qiskit.},label={listing:bellqiskit},captionpos=b,language=Python]
no_qubits = 2

circuit = QuantumCircuit(no_qubits)
circuit.h(0)
circuit.cnot(0, 1)

backend = Aer.get_backend("unitary_simulator")
job = execute(circuit, backend)
\end{lstlisting}

% \begin{figure}[htbp!] 
%     \centering    
%     \includegraphics[width=0.7\textwidth]{UML_unitary.png}
%     \caption[Unitary matrix simulator (UML)]{UML diagram of unitary matrix simulators (standard and accelerated).}
%     \label{fig:umlunitary}
% \end{figure}

\begin{figure}[htbp!] 
    \centering    
    \includegraphics[width=0.6\textwidth]{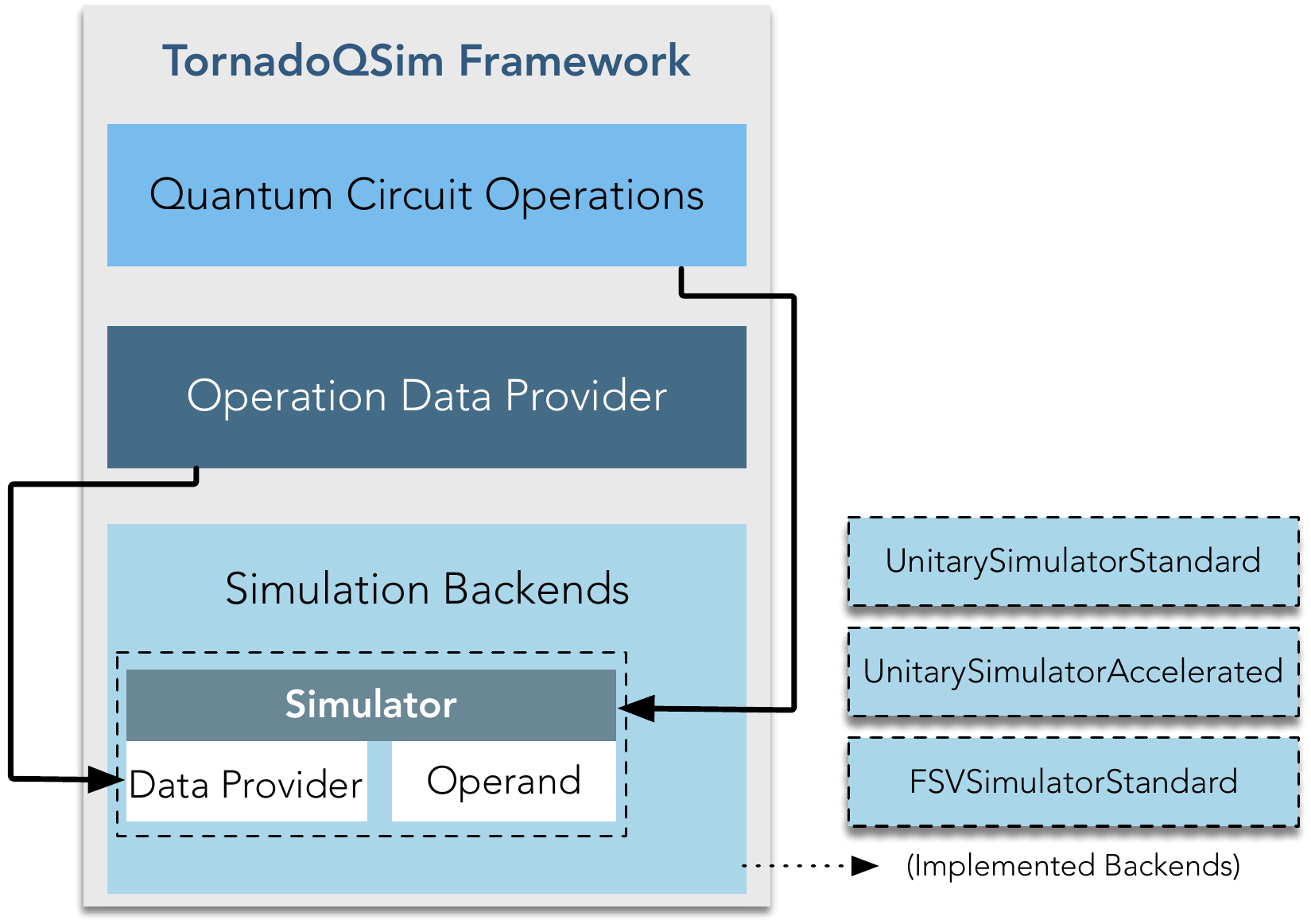}
    \caption{The expandability of backends in TornadoQSim.}
    \label{fig:expandable_backends}
\end{figure}

% \subsubsection{Command Line Interface}
% Users can configure the TornadoQSim quantum simulator by declaring information, such as a circuit type, a number of qubits, and a backend of preference within the submitted requests.
% Furthermore, users can define new circuits directly in Java by specifying the qubits on which a particular quantum operation should be performed, which is similar to other state-of-the-art quantum simulators (e.g.,\ Qiskit~\cite{qiskit_docs2021}).

\subsection{Modularity \& Expandability}
\label{sec:modularity}
To demonstrate the expandability of TornadoQSim, this section discusses the add-on implementation of a full-state vector simulation backend.
In general, every new simulation backend should be composed of at least the following three main components, as illustrated in Figure~\ref{fig:expandable_backends}. 
First, a custom data provider class takes the raw quantum gate data from the \textit{Operation Data Provider} and implements supporting methods to transform the data into a format used by the operand class and the simulator class.
Next, an operand class provides a potentially accelerated implementation of main mathematical operations required for the specific simulation model.
Lastly, a simulator class implements the \textit{Simulator} interface (\ref{sec::simulator_backends}) and uses the above mentioned classes to perform the actual simulation process (Section~\ref{sec::simulation_flow}).

Specifically, the full-state vector simulation backend focuses on an iterative way of simulating the provided quantum circuit, rather than constructing large unitary matrices for each step. 
The backend is composed of the following classes: \textit{FsvDataProvider}, \textit{FsvOperand}, and \textit{FsvSimulatorStandard}.
The \textit{FsvDataProvider} class does not do any transformation to the raw gate data and provides the data as they are (unitary matrices).
This is because no specific transformation is required by this simulation backend. 
The \textit{FsvOperand} class implements two methods (applyGate and applyControlGate), which apply the gates by iterating over the full state vector.
Finally, \textit{FsvSimulatorStandard} implements the \textit{Simulator} interface and orchestrates the simulation process of the supplied quantum circuit.

% by users who want to test new simulation backends or quantum circuits

\subsection{Transparent Hardware Acceleration}
\label{sec:system_hw_accel}
An additional point that differentiates TornadoQSim from other Java quantum simulators (e.g.,\ Strange) is that 
it can leverage JIT complilation to offload parts of its simulation backends on heterogeneous co-processors.
This is an advantage of TornadoQSim, as it is not restricted to use specific pre-compiled kernels, but it can potentially produce parallel implementations of additional simulation backends written in Java.

To produce a parallel implementation of these mathematical calculations for the \textit{UnitarySimulatorAccelerated} backend,
we utilized the \texttt{@Parallel} annotation, as was introduced in Section~\ref{sec:hardware_acceleration_tornadovm}.
Listing~\ref{listing:tornadovm-mxm} shows how we used the annotation in lines 2 and 3 to define a parallel implementation of the matrix multiplication operation across all steps (stage \circled{5} in Figure~\ref{fig:simulation-flow}).
The exclusion of the annotation produces a functionally equivalent single-threaded implementation as it is performed in the \textit{UnitarySimulatorStandard} backend.
Therefore, TornadoQSim provides both standard and accelerated unitary matrix simulation backends as a means to assess the seamless contribution of heterogeneous hardware devices to the acceleration of the simulation time.

\begin{lstlisting}[caption={TornadoVM implementation of the matrix multiplication in TornadoQSim.},label={listing:tornadovm-mxm},captionpos=b,language=Java]
protected static void matrixMultiplication(float[] realA, float[] imagA, final int rowsA, final int colsA, float[] realB, float[] imagB, final int colsB, float[] realC, float[] imagC) {
  for (@Parallel int i = 0; i < rowsA; i++) {
    for (@Parallel int j = 0; j < colsB; j++) {
      int indexC = (i * rowsA) + j;
      realC[indexC] = 0;
      imagC[indexC] = 0;
      for (int k = 0; k < colsA; k++) {
        int indexA = (i * colsA) + k;
        int indexB = (k * colsB) + j;
        realC[indexC] += (realA[indexA] * realB[indexB]) - (imagA[indexA] * imagB[indexB]);
        imagC[indexC] += (realA[indexA] * imagB[indexB]) + (imagA[indexA] * realB[indexB]);
      }
    }
  }
}
\end{lstlisting}

\section{Experimental Evaluation}
\label{section::evaluation}
This section presents the performance assessment of TornadoQSim against a vanilla implementation in Java (hereafter named as \textit{Vanilla}) and the \textit{Qiskit} simulator.
The first comparison aims to show the benefits of employing transparent JIT compilation for heterogeneous accelerators through TornadoVM, 
while the second comparison scopes the overall comparison against a state-of-the-art quantum simulator.
Our experiments are executed following the same methodology (Section~\ref{sec::methodology})
Section~\ref{sec::evaluated_circuits} presents the evaluated circuits, while Section~\ref{sec::results_vs_java} discusses the performance evaluation of TornadoQSim on GPUs against the functionally equivalent Java implementation.
Finally, Section~\ref{sec::results_vs_qiskit} presents the performance assessment of TornadoQSim against Qiskit.

\subsection{Experimental Methodology}
\label{sec::methodology}
All experiments have been performed on a server that contains an Nvidia GP100 GPU, using the same operating system and heap size (Table~\ref{table::platform_characteristics}) to simulate all quantum circuits presented in Table~\ref{table::bench-circuits}.
We performed the warm-up process which included at least 40\footnote{This number of iterations has been sufficient to ensure the hotness of the meassured code segments.} executions prior to the actual timing of all systems, in order to warm up the JVM and fairly compare all frameworks.
The reported results are the average of the next 11 iterations of each measurement.
Note that we have validated that the deviations across the obtained measurements are insignificant.

\begin{table}[t!]
	\centering
	\caption{The experimental hardware and software characteristics of the testbed.}
	\label{table::platform_characteristics}
	\begin{tabular}{|l|c|}
		\hline
        \textbf{CPU}  & Intel Core i7-7700K CPU @ 4.20GHz \\ \hline
		\textbf{Memory} &  64 GB \\ \hline
		\textbf{JVM} & OpenJDK 17 64-bits \\ \hline
		\textbf{JVM Heap Size} & 16 GB \\ \hline
		\textbf{GPUs} & Nvidia Quadro GP100 GPU \\ \hline
        \textbf{GPU Memory (DRAM)} & 16 GB \\ \hline
		\textbf{Operating System} & CentOS Linux release 7.9.2009 \\ \hline
		\textbf{TornadoVM Backends} & OpenCL \& PTX \\ \hline
	\end{tabular}
	% \vspace{-1.5em}
\end{table}

\begin{table*}[h]
	\centering
	\caption{Quantum circuits used for the benchmarking of the simulators.}
	\label{table::bench-circuits}
	\begin{tabular}{|c|c|c|c|c|c|c|c|}
        \hline
		\textbf{Quantum Circuits}   & \textbf{Number of Qubits} \\ \hline\hline
        \textbf{Quantum Fourier Transform (QFT)} & 4-12 \\ \hline
		\textbf{Entanglement} & 4-12 \\ \hline
		\textbf{DeutschJozsa} & 4-12 \\ \hline
	\end{tabular}
	% \vspace{-1em}
\end{table*}

\subsection{Evaluated Quantum Circuits}
\label{sec::evaluated_circuits}
To assess the performance comparison of TornadoQSim, we analyzed the performance of the accelerated implementation of \textit{UnitarySimulatorStandard} (presented in Section~\ref{sec::simulation_interfaces}) against the non-accelerated implementation (i.e.,\ Vanilla) for three quantum circuits that were previously described in Section~\ref{sec:quantum_programs}: Fully Entangled Circuit, Deutsch-Jozsa Algorithm, Quantum Fourier Transform.
Some of these circuits are used for the evaluation of other quantum simulators~\cite{quest2019, projectq2017, projectq2018}.
They compose a suitable set of circuits which demostrates different characteristics of quantum applications in the quantum computing field.
For instance, the Fully Entangled Circuit can demonstrate the entanglement property.
Additionally, the Deutsch-Jozsa Algorithm is used as an example to showcase the advantage of quantum computing for a decision problem,
while Quantum Fourier Transform acts as a base for many other practical algorithms (e.g.,\ the Shor's algorithm) in the quantum computing field.

Table~\ref{table::bench-circuits} presents the number of qubits that we evaluated for all circuits. 
Each circuit is simulated for different number of qubits ranging from 4 to 12 qubits.
Note that the performance evaluation of the non accelerated implementation of the \textit{UnitarySimulatorStandard} backend for 12 qubits was impactical, due to a prolonged execution time.
Thus, the experiments presented in Section~\ref{sec::results_vs_java} run all circuits with up to 11 qubits.
On the contrary, the experiments in Section~\ref{sec::results_vs_qiskit} that compare the performance of the accelerated simulation backend against Qiskit run with up to 12 qubits.

\subsection{Performance Evaluation on GPUs}
\label{sec::results_vs_java}
In the conducted experiments we assessed the performance of three accelerated implementations for each quantum circuit; one implementation for a multicore CPU (\textit{CPU-OpenCL}), and two implementations for a GPU (\textit{GPU-OpenCL}, and \textit{GPU-PTX}).
All the accelerated implementations leverage the TornadoVM JIT compiler for seamless hardware acceleration.
The former two implementations consume hardware kernels generated by the OpenCL backend of TornadoVM, while the latter consumes hardware kernels generated by the TornadoVM's PTX backend which runs exclusively on Nvidia GPUs.
The configuration of the computer system used to run our experiments is listed in Table~\ref{table::platform_characteristics}.
We run our experiments following the experimental methodology described in Section~\ref{sec::methodology}.

\begin{figure}[t!]
	\centering
    \includegraphics[width=0.7\textwidth]{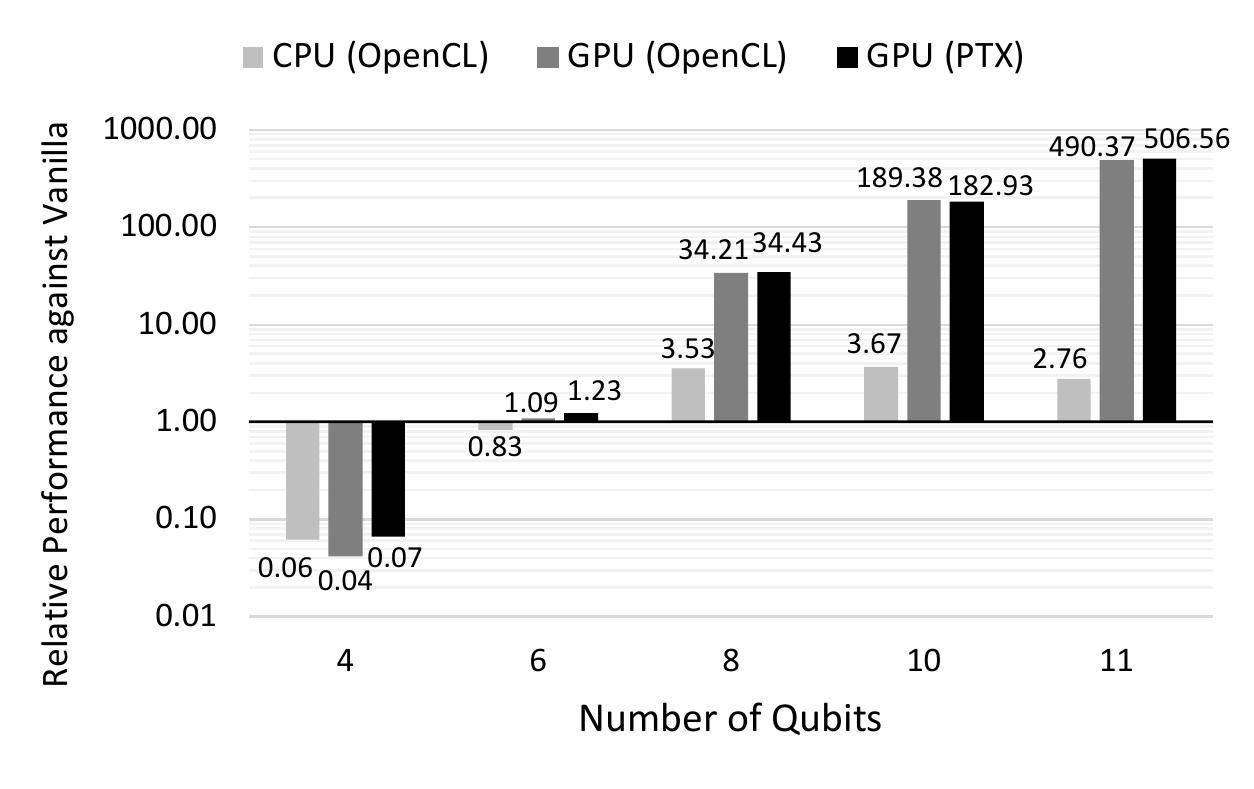}
    \caption{Comparative relative end-to-end performance evaluation of TornadoQSim for a Fully Entangled circuit. The implementations on three different configurations (CPU-OpenCL, GPU-OpenCL, GPU-PTX) are presented against the Vanilla implementation. The higher, the better.}
    \label{fig:perf-entanglement}
	% \vspace{-1em}
\end{figure}

\paragraph{Evaluation of a Fully Entangled Quantum Circuit}
Figure~\ref{fig:perf-entanglement} shows the relative end-to-end performance of the accelerated implementations executed on a multi-core CPU, GPU (OpenCL) and GPU (CUDA-PTX), against the non-accelerated simulation (Vanilla). 
The obtained results show that the implementations that run on heterogeneous hardware are outperformed by the \textit{Vanilla} implementation for small quantum circuits up to 6 qubits. 
The reason is that the computation time is not significant and therefore the overall performance is penalized by the time spent for packaging data in the data structures of the utilized programming model (OpenCL, CUDA in case of PTX) and
sending data to the GPUs via the PCIe interconnect.
However, for large circuits that deploy more qubits, a significant performance increase is achieved. 
The highest performance speedup, 506.5$x$, is achieved when the simulation run on the GPU with OpenCL driver for a 11-qubit circuit.

\begin{figure}[t!]
	\centering
    \includegraphics[width=0.7\textwidth]{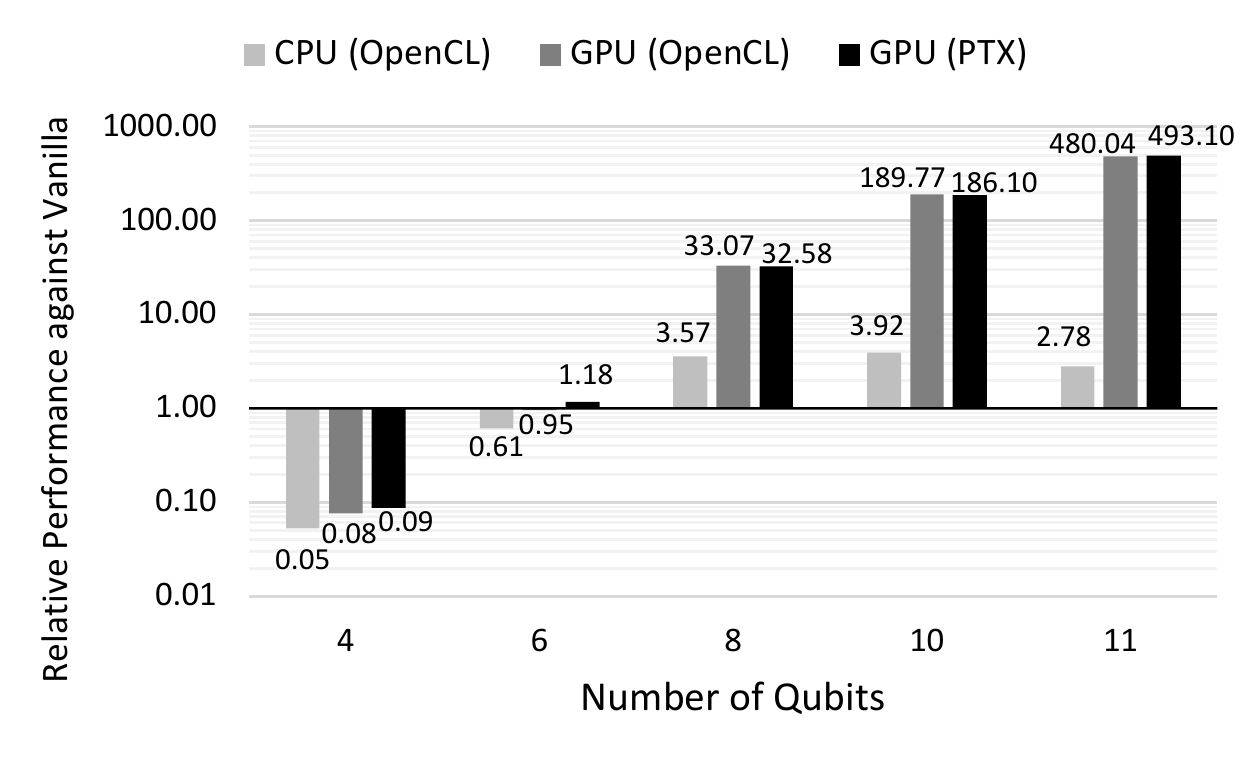}
    \caption{Comparative relative end-to-end performance evaluation of TornadoQSim for a Deutsch-Jozsa quantum algorithm. The implementations on three different configurations (CPU-OpenCL, GPU-OpenCL, GPU-PTX) are presented against the Vanilla implementation. The higher, the better.}
    \label{fig:perf-deutch}
	% \vspace{-1em}
\end{figure}

\paragraph{Evaluation of a Deutsch-Jozsa Quantum Algorithm}
Figure~\ref{fig:perf-deutch} shows the relative end-to-end performance of the accelerated against the non-accelerated (\textit{Vanilla}) simulations for the Deutsch-Jozsa Quantum Algorithm.
The results follow the trend of the previous experiments, showing a performance slowdown for small quantum circuits up to 6 qubits, but a significant performance speedup for larger circuits.
Both GPU implementations (OpenCL and PTX) show competitive performance for the simulations with different number of qubits.
The highest performance speedup is observed for the simulation of a circuit with 11 qubits, in which the GPU (PTX) implementation outperforms the \textit{Vanilla} implementation by 493.10$x$.

\begin{figure}[t!]
	\centering
    \includegraphics[width=0.7\textwidth]{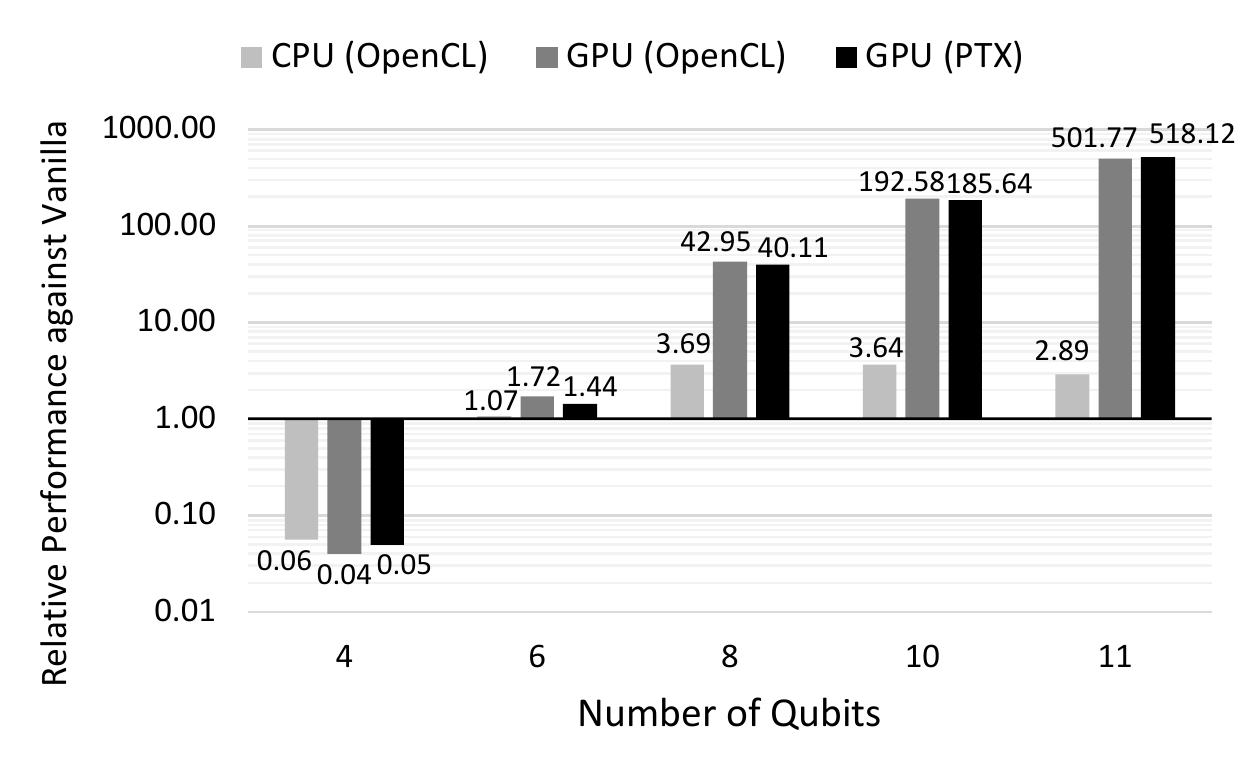}
    \caption{Comparative relative end-to-end performance evaluation of TornadoQSim for a Quantum Fourier Transform algorithm. The implementations on three different configurations (CPU-OpenCL, GPU-OpenCL, GPU-PTX) are presented against the Vanilla implementation. The higher, the better.}
    \label{fig:perf-qft}
	% \vspace{-1em}
\end{figure}

\paragraph{Evaluation of a Quantum Fourier Transform Algorithm}
The final circuit that we evaluated is a Quantum Fourier Transform algorithm.
Figure~\ref{fig:perf-qft} illustrates the relative end-to-end performance of the accelerated implementations against the non-accelerated simulation (Vanilla) in logarithmic scale. 
Similarly with the previously evaluated circuits, the \textit{Vanilla} implementation outperfoms the implementations that utilize CPU and GPU offloading for small quantum circuits up to 6 qubits. 
However, for large circuits that deploy more qubits, the accelerated implementations show a performance speedup of up to 518.12$x$ (GPU OpenCL for 11 qubits).

\begin{figure}[ht] 
% \begin{subfigure}[a]{0.5\textwidth}
% 	\includegraphics[width=\linewidth]{figures/qiskit-perf-entaglement}
% 	\caption{Fully Entangled Quantum Circuit.}
% 	\label{fig:qiskit-entanglement}
%   \end{subfigure}
\begin{subfigure}[a]{0.5\textwidth}
	\includegraphics[width=\linewidth]{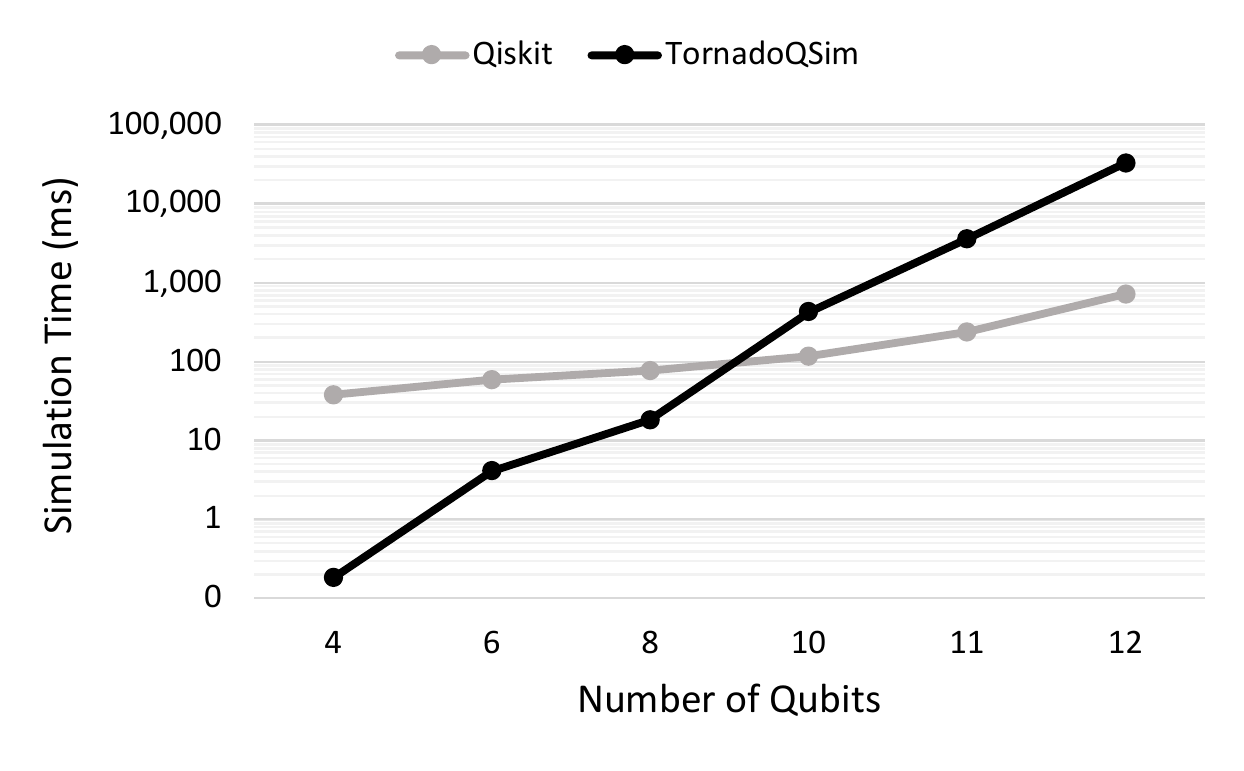}
	\caption{Deutsch-Jozsa Quantum Algorithm.}
	\label{fig:qiskit-deutch}
\end{subfigure}
\begin{subfigure}[a]{0.5\textwidth}
	\centering
	\includegraphics[width=\linewidth]{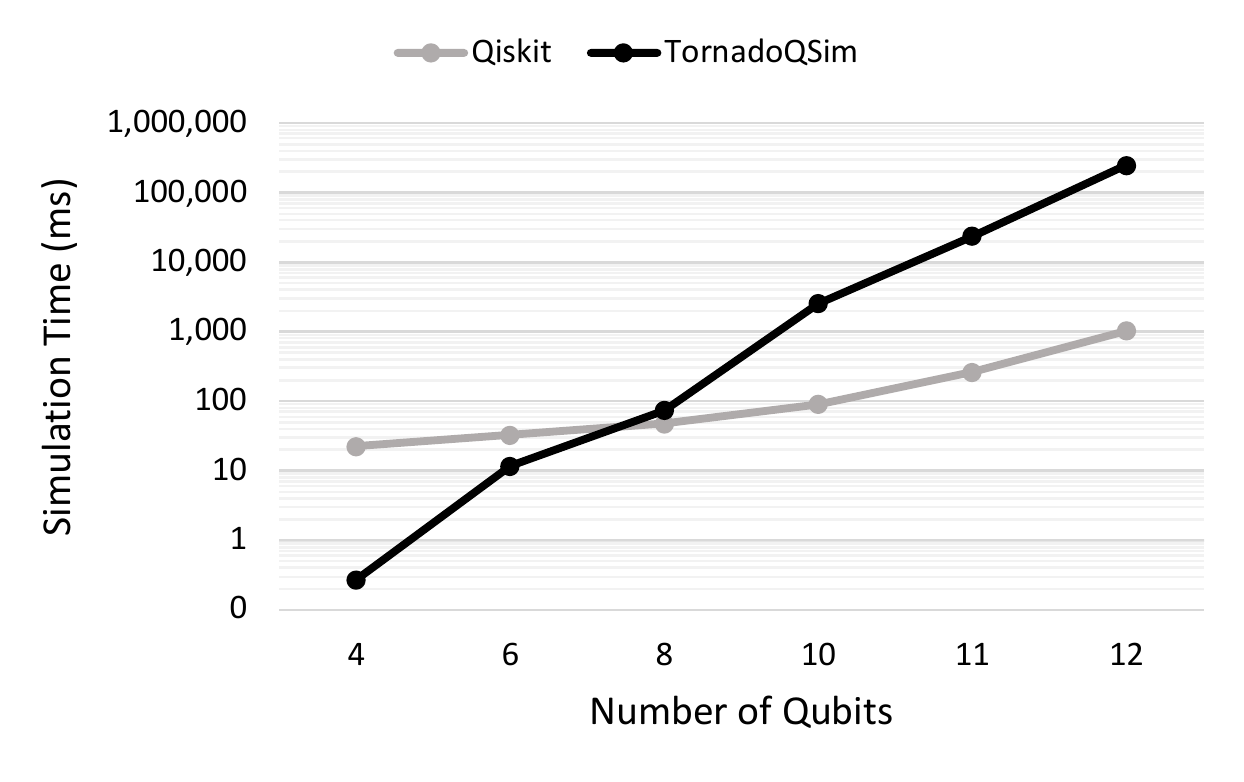}
	\caption{Quantum Fourier Transform Algorithm.}
	\label{fig:qiskit-qft}
\end{subfigure}
\caption{End-to-end simulation time of the best TornadoQSim implementation against Qiskit. The lower, the better.}\label{fig:qiskit_perf}
\end{figure}

\subsection{Performance Evaluation against Qiskit}
\label{sec::results_vs_qiskit}
This paragraph aims to analyze the performance of TornadoQSim against a state-of-the-art quantum simulator; Qiskit\cite{qiskit_docs2021}.
The comparison with Qiskit (rather than other frameworks) is performed mainly because Qiskit provides a unitary matrix simulation back-end, thus enabling a fair comparison. 
Other frameworks, such as Strange~\cite{strange_docs2021}, typically use more efficient simulation method.

The best performing implementation of TornadoQSim is compared with Qiskit, which is configured to use a unitary simulator as the simulation backend. 
This means that for small circuits, TornadoQSim, is used without the acceleration (Vanilla) of the unitary matrix simulation backend, while for large circuits\footnote{In this section, we include performance numbers of TornadoQSim for simulation of 12 qubits via the hardware accelerated backend.} the accelerated simulation backend is employed (GPU-OpenCL or GPU-PTX).
The non-accelerated setting of the low-level unitary simulator backend is used for Qiskit (C++, without the use of OpenMP or other acceleration).
The comparison is performed when running the Deutsch-Jozsa algorithm and the QFT algorithm for several circuit sizes; ranging from 4 qubits to 12 qubits.
% The performance graph (Figure~\ref{fig:tornadoperf}), where time required to run the particular circuit is plotted. The less time/memory required to simulate the circuit, the better. Note that the y axis is in logarithmic scale. Also, the comparison with Qiskit (rather than other frameworks) is performed mainly because Qiskit provides a unitary matrix simulation beach-end, thus enabling a fair comparison. Other frameworks, such as Strange, typically use the more efficient state vector simulation method.
Figure~\ref{fig:qiskit_perf} presents the simulation time of both algorithms for the TornadoQSim (black line) and Qiskit (grey line). 
The less time required to simulate the circuit, the better.
Note that the vertical axis is reported in logarithmic scale. 

Figure~\ref{fig:qiskit-deutch} shows that for small circuits (up to 4 qubits) the \textit{Vanilla} implementation of TornadoQSim is faster than Qiskit by 208$x$.
Then for medium circuits (6 to 8 qubits), the GPU accelerated implementation of TornadoQSim outperforms Qiskit by up to 4$x$.
In large circuits (more than 8 qubits), TornadoQSim is slower for an order of magnitude.
Similar performance trend is also observed for the QFT algorithm in Figure~\ref{fig:qiskit-qft}.
In this case, the \textit{Vanilla} implementation of TornadoQSim is 82$x$ faster for small circuits (up to 4 qubits).
For 6 qubits, the GPU accelerated implementation of TornadoQSim outperforms Qiskit by up to 2.8$x$,
while for 8 qubits, it is 46\% slower than Qiskit.
Finally, Qiskit outperforms TornadoQSim by an order of magnitude for simulations of circuits that have more than 8 qubits.

At this stage, note that the performance assessment of both systems (TornadoQSim and Qiskit) is performed with regards to the end-to-end simulation time.
This means that the simulation via TornadoQSim is executed on top of the Java Virtual Machine (JVM), while via Qiskit it runs natively as it is implemented in C++.
A close inspection of the performance difference between the two simulators is out of the scope of this paper, as it has to be performed at the system's level.
In this work, we rather aim to show that the proposed simulator is an open-source framework that offers a modular design that can be easily extended by users for adding further simulation functionality (new simulation circuits or backends); and it is functionally correct.
\section{Conclusions}
This paper presents TornadoQSim, an open-source quantum simulation framework implemented in Java.
The proposed system can be used as an open framework for programmers who want to simulate quantum circuits from Java, while also capitalizing the performance benefits of hardware acceleration at no further programming cost.
The proposed framework combines a modular and easily expandable architecture that enables users to add customized simulation circuits and simulation backends.
TornadoQSim is tightly coupled with TornadoVM to automatically offload any computationally intensive parts of the simulation backends on heterogeneous hardware accelerators.
The performance evaluation of TornadoQSim was conducted on a variety of hardware (multi-core CPU and GPU) and showed that small circuits are simulated faster in Java,
while for large circuits (more than 6 qubits) the simulation is accelerated by 506.5$x$, 493.10$x$ and 518.12$x$ for a fully entangled circuit, a Deutsch-Jozsa quantum algorithm, and a Quantum Fourier Transform algorithm, respectively.
Additionally, we performed a performance comparison of the best TornadoQSim implementation of unitary matrix against a semantically equivalent simulation via Qiskit.
The comparative evaluation showed that the simulation with TornadoQSim is faster for small circuits, while for large circuits Qiskit outperforms TornadoQSim by an order of magnitude.

As future work, we plan to analyze the source of slow simulation times for large circuits with the aim of achieving competitive performance against other state-of-the-art quantum simulators.
Finally, we plan to increase the maturity of the simulator by expanding the number quantum circuits and simulation backends as well as enabling hardware acceleration for the full-state vector backend.
% and share task schedules among layers that could optimize any redundant copies of the current implementation.

\begin{acks}
    This work is partially funded by grants from Intel Corporation and the European Union Horizon 2020 ELEGANT 957286. 
    Additionally, this work is supported by the Horizon Europe AERO, INCODE, ENCRYPT and TANGO projects which are funded by UKRI grant numbers 10048318, 10048316, 10039809 and 10039107.
\end{acks}

\appendix
\section{Elaborate Analysis of Quantum Systems}
\label{sec::appendix-analysis-quantum}

\subsection{Additional Simulation Backends}
\subsubsection{Full State Vector Backends}
\label{sec::appendix-fsv-tensor}

To simulate a quantum circuit using the full state vector technique, it is necessary to iterate through the state vector for each quantum gate and apply the update equation (i.e.,\ linear map) to the qubits related to the quantum gate.
Equation~\ref{eq12} presents the mathematical description of this optimization technique for a single qubit gate.

\[
U = 
\begin{bmatrix}
    U_a & U_b \\
    U_c & U_d
\end{bmatrix}
\]

\begin{equation}\label{eq12}
\begin{split}
    \alpha_{...0_i...} = U_a \cdot \alpha_{...0_i...} + U_b \cdot \alpha_{...1_i...} \\
    \alpha_{...1_i...} = U_c \cdot \alpha_{...0_i...} + U_d \cdot \alpha_{...1_i...}
\end{split}
\end{equation}

A similar formula is used for two-qubit gates.
As the unitary matrix is not required to be constructed for each step, the required space is significantly decreased.
For instance, to simulate the same circuit of 20 qubits with this method, it would be required to store only the full state vector ($2^{20}$ probability amplitudes).
Considering a 64-bit representation for complex numbers, it would result in around 8.4 MB of memory space, which is notably less than 8.8 TB (needed for the unitary matrix approach).
A simulator using this method is typically named as a full state vector simulator~\cite{hidary2019, kelly2018}.

\subsubsection{Tensor Network Backends}
\begin{figure}[htbp!] 
    \centering    
    \includegraphics[width=0.9\textwidth]{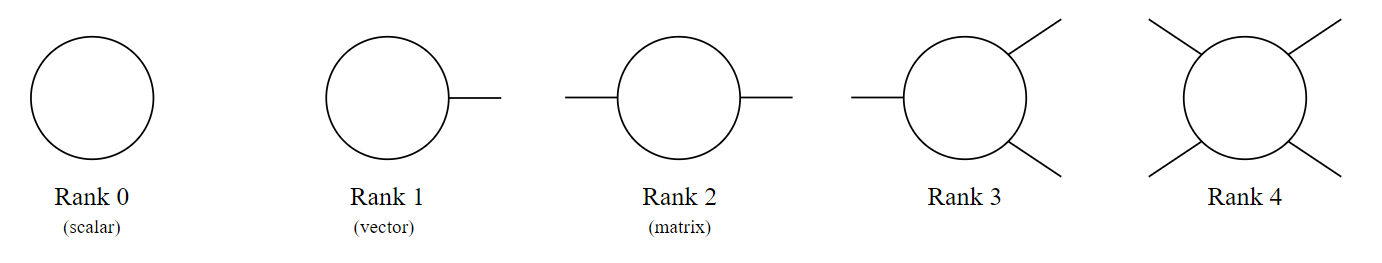}
    \caption{Tensor representation in tensor networks.}
    \label{fig:tn}
\end{figure}

An alternative approach is to represent the quantum circuit by using a tensor network.
Tensor network is a mathematical model of quantum many-body states~\cite{viamontes2003}.
In essence, every quantum circuit corresponds to a tensor network.
A tensor is a mathematical object with rank $n$, where every entry is indexed by $n$ indices~\cite{iqs2020}.
Several examples of tensors with different ranks exist, such as a scalar (rank 0), a vector (rank 1) and a matrix (rank 2).
Figure~\ref{fig:tn} presents a graphical representation of tensors with various ranks.
In the context of quantum circuits, a rank 1 tensor is used for a single qubit state, while a rank 2 tensor represents a single qubit quantum gate (two states).
Subsequently, a two-qubit quantum gate can be represented by a rank 4 tensor.

Tensors are interconnected with edges, where each edge corresponds to a particular index of the tensor~\cite{iqs2020}.
In case that two tensors share a common index (are connected via an edge), a contraction operation can be performed to combine the tensors together.
The contraction operation is defined as Einstein summation over the shared index,
and the objective is to efficiently contract the whole network into a state vector.
Alternatively, a tensor with a large rank can be split into several tensors with smaller ranks, by using an operation called Singular Value Decomposition (SVD).
To simulate a quantum circuit using tensor networks means to perform a contraction operation across the whole network.
However, the order of operations matters as the complexity of contraction depends on the maximum rank of the tensors involved in the operation.
The efficiency of the quantum circuit simulation using the tensor network approach can be improved with a correct contraction strategy.
In other words, tensors should be contracted in a particular order to minimize the complexity~\cite{iqs2020}.

\subsubsection{Graph Based Backends}
\begin{figure}[htbp!] 
    \centering    
    \includegraphics[width=0.3\textwidth]{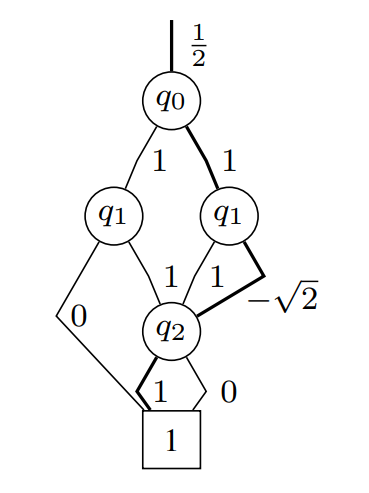}
    \caption[Graph representation of the quantum state]{Representation of the quantum state using a quantum decision diagram (Adapted from~\cite{zulehner2018}).}
    \label{fig:graph}
\end{figure}

Other methods have emerged to simulate quantum circuits based on quantum decision diagrams,
such as Quantum Information Decision Diagrams (QuiDD)~\cite{viamontes2003} and Quantum Multi-Valued Decision Diagrams (QMDD)\cite{zulehner2018}.
The key idea of these methods is the decomposition of the state vectors in subvectors, till the subvectors reach a granularity of a single element that corresponds to a complex number~\cite{zulehner2018}.
In this case, the state vector can be represented by a decision diagram (i.e.,\ a directed acyclic graph), as shown in Figure~\ref{fig:graph}.
Figure~\ref{fig:graph} shows an example of a quantum decision diagram that is used as a structure for a quantum state $\ket{\psi}$~\cite{zulehner2018}.
Each probability amplitude in a quantum decision diagram is calculated as a product of the edge values.
Therefore, the left edge that departs from a node (qubit) corresponds to a qubit with quantum state equal to 0,
whereas the right edge that departs from a node (qubit) corresponds to a qubit with quantum state equal to 1.
For a three qubit quantum system (Figure~\ref{fig:graph}), the individual probability amplitudes can be calculated as shown in Equation~\ref{eq13}.

\begin{equation}\label{eq13}
    \ket{\psi} = 
    \begin{bmatrix}
        \alpha_{000} \\
        \alpha_{001} \\
        \alpha_{010} \\
        \alpha_{011} \\
        \alpha_{100} \\
        \alpha_{101} \\
        \alpha_{110} \\
        \alpha_{111} 
    \end{bmatrix}
    =
    \begin{bmatrix}
        0 \\
        0 \\
        \frac{1}{2} \\
        0 \\
        \frac{1}{2} \\
        0 \\
        \frac{-1}{\sqrt{2}} \\
        0
    \end{bmatrix}
    =
    \begin{bmatrix}
        \frac{1}{2} \cdot 1 \cdot 0 \\
        \frac{1}{2} \cdot 1 \cdot 0 \\
        \frac{1}{2} \cdot 1 \cdot 1 \cdot 1 \\
        \frac{1}{2} \cdot 1 \cdot 1 \cdot 0 \\
        \frac{1}{2} \cdot 1 \cdot 1 \cdot 1 \\
        \frac{1}{2} \cdot 1 \cdot 1 \cdot 0 \\
        \frac{1}{2} \cdot 1 \cdot -\sqrt{2} \cdot 1 \\
        \frac{1}{2} \cdot 1 \cdot -\sqrt{2} \cdot 0 
    \end{bmatrix}
\end{equation}

\bibliographystyle{ACM-Reference-Format}
\bibliography{refs}

\end{document}